\documentclass[12pt,preprint]{emulateapj}

\usepackage{amssymb}
\usepackage{amsmath}
\usepackage{lscape}
\usepackage{subfigure}

\newcommand{\ssize}{67}
\newcommand{\ssizeabdor}{31}
\newcommand{\ssizebpic}{11}
\newcommand{\ssizeherlyr}{7}
\newcommand{\ssizetuchor}{2}
\newcommand{\ssizetwa}{16}

\newcommand{\nruns}{twelve}
\newcommand{\ms}{$M_\odot$}
\newcommand{\mj}{$M_{\rm J}$}

\shorttitle{Mapping the Shores of the Brown Dwarf Desert III}
\shortauthors{Evans et al}

\begin{document}

\title{Mapping the Shores of the Brown Dwarf Desert III: Young Moving Groups}

\author{Evans, T.M.\altaffilmark{1,2}, Ireland,
  M.J.\altaffilmark{1,3,4}, Kraus, A.L. \altaffilmark{5,9}, 
Martinache, F. \altaffilmark{6}, Stewart, P.\altaffilmark{1}, 
Tuthill, P.G.\altaffilmark{1}, Lacour, S. \altaffilmark{7,1},
Carpenter, J.M. \altaffilmark{8} and Hillenbrand, L.A. \altaffilmark{8}}
\altaffiltext{1}{Sydney Institute for Astronomy (SIfA), 
  School of Physics, University of Sydney NSW
  2006, Australia}
\altaffiltext{2}{Department of Physics, Denys Wilkinson Building, Keble
  Road, Oxford, OX1 3RH, UK: tom.evans@astro.ox.ac.uk}
\altaffiltext{3}{Department of Physics and Astronomy, Macquarie
  University, NSW 2109, Australia}
\altaffiltext{4}{Australian Astronomical Observatory, PO Box 296, Epping, NSW 1710, Australia}
\altaffiltext{5}{Institute for Astronomy, University of Hawaii, 2680 Woodlawn
  Drive, Honolulu, HI 96822, USA}
\altaffiltext{6}{National Astronomical Observatory of Japan,
  Subaru Telescope, Hilo, HI 96720, USA}
\altaffiltext{7}{Observatoire de Paris, LESIA, CNRS/UMR 8109, 92190 Meudon, France}
\altaffiltext{8}{Department of Astrophysics, California Institute of Technology, 
  MC 105-24, Pasadena, CA 91125}
\altaffiltext{9}{Hubble Fellow}

\begin{abstract}
We present the results of an aperture masking interferometry survey for substellar companions around \ssize \ members of the young ($\sim$\,8--200\,Myr) nearby ($\sim$\,5--86\,pc) AB~Doradus, $\beta$~Pictoris, Hercules-Lyra, TW~Hya, and Tucana-Horologium stellar associations. Observations were made at near infrared wavelengths between 1.2--3.8\,$\mu$m using the adaptive optics facilities of the Keck~II, VLT UT4, and Palomar Hale Telescopes. Typical contrast ratios of $\sim$\,100--200 were achieved at angular separations between $\sim$\,40--320\,mas, with our survey being 100\% complete for companions with masses below $\sim$0.25\,\ms \ across this range. We report the discovery of a $0.52 \pm 0.09$\,\ms \ companion to HIP\,14807, as well as the detections and orbits of previously known stellar companions to HD\,16760, HD\,113449, and HD\,160934. We show that the companion to HD\,16760 is in a face-on orbit, resulting in an upward revision of its mass from $M_2 \sin i \sim 14$\,\mj \ to $M_2 = 0.28 \pm 0.04$\,\ms. No substellar companions were detected around any of our sample members, despite our ability to detect companions with masses below 80\,\mj \ for 50 of our targets: of these, our sensitivity extended down to 40\,\mj \ around 30 targets, with a subset of 22 subject to the still more stringent limit of 20\,\mj. A statistical analysis of our non-detection of substellar companions allows us to place constraints on their frequency around $\sim$\,0.2--1.5\,\ms \ stars. In particular, considering companion mass distributions that have been proposed in the literature, we obtain an upper limit estimate of $\sim$\,9--11\% for the frequency of 20--80\,$M_{\rm J}$ companions between 3--30\,AU at 95\% confidence, assuming that their semimajor axes are distributed according to $d\mathcal{N}/da \propto a^{-1}$ in this range. 
\end{abstract}


\section{Introduction}

In the past few years, direct imaging surveys have begun to build up a picture of the mass and semimajor axis distributions of substellar companions at separations beyond $\sim$\,20--30\,AU \citep[eg.][]{2007ApJS..173..143B, 2006AJ....132.1146C, 2010A&A...509A..52C, 2007A&A...472..321K, 2007ApJ...670.1367L, 2005AJ....130.1845L, 2005ApJ...625.1004M, 2009ApJS..181...62M}. Meanwhile, statistical analyses of radial velocity results have tended to focus on objects with masses below $\sim$\,10$M_{\rm J}$ out to separations of $\sim$\,3\,AU \citep{2008PASP..120..531C, 2010Sci...330..653H}. However, given the observational biases of radial velocity and direct imaging surveys, the separation range of $\sim$\,3--30\,AU has remained relatively unexplored. 

Aperture masking interferometry is a direct detection technique that is well suited for detecting substellar companions with masses of $\sim$\,10\,$M_{\rm J}$ and semimajor axes within $\sim$\,30\,AU around young, nearby stars. For instance, it has been used to conduct surveys for substellar companions around members of the Upper Scorpius \citep{2008ApJ...679..762K} and Taurus-Auriga \citep{2011ApJ...731....8K} associations, as well as to measure the dynamical mass of the brown dwarf companion to GJ~802b \citep{2008ApJ...678..463I}, show that CoKu Tau/4 is a binary system rather than a transitional disk \citep{2008ApJ...678L..59I}, and place limits on possible companions existing within 10\,AU of HR\,8799 \citep{2011ApJ...730L..21H}. Recently, the technique has also produced the first direct detection of a young exoplanet still undergoing formation within the transitional disk of LkCa15 (Kraus \& Ireland, submitted) and a similar detection of an object within the gap of the T Cha disk \citep{2011A&A...528L...7H}.

This paper presents the results of an aperture masking survey of \ssize \ members of the AB~Doradus (AB~Dor), $\beta$~Pictoris ($\beta$~Pic), Hercules-Lyra (Her-Lyr), Tucana-Horologium (Tuc-Hor), and TW~Hya (TWA) moving groups. At least 49 of our targets have been observed previously as part of deep imaging surveys, but these observations have typically been sensitive to different orbital separations than those that are probed here. We chose our targets based on their youth (8--200\,Myr) and proximity (5--86\,pc). The former ensures that any substellar companions are still glowing relatively brightly at infrared wavelengths following their recent formation, while the latter allows smaller absolute separations to be explored for a given angular separation. 

The paper is organized as follows. In Section~\ref{sec:am}, we provide a brief overview of the aperture masking technique. In Section~\ref{sec:sample}, we describe our survey sample. In Section~\ref{sec:obsdatared} we summarize the observations that were made and how the data were reduced. In Section~\ref{sec:binaryfitting}, we explain how we searched for companions to the target stars in the reduced data and how we derived the survey detection limits. In Section~\ref{sec:results} we report our results, including the detections and orbits for stellar companions around HIP\,14807, HD\,16760, HD\,113449, and HD\,160934. However, no substellar companions were detected, and in Section~\ref{sec:analysis} we present a statistical analysis of this null result before concluding in Section~\ref{sec:conclusion}.

\section{Aperture Masking} \label{sec:am}

The aperture masking technique works by placing an opaque, perforated mask at or near the pupil plane of a telescope (\citealt{fizeau1868}; \citealt{1891Natur..45..160M}; more recently, see \citealt{2000PASP..112..555T, 2006SPIE.6272E.103T, 2010SPIE.7735E..56T}). This converts the single aperture into a multi-element interferometer. Each pair of holes in the mask acts as an interferometric baseline, resulting in an interferogram being projected onto the image plane. 

The complex visibility~$V$ \citep{michelson1890} of the source brightness distribution~$S$ is sampled by taking the 2D Fourier transform of the measured interferogram~$I$. This follows from the Van Cittert-Zernike Theorem, which states that the normalized complex visibility is equal to the Fourier transform of the brightness distribution:
\begin{eqnarray}
 V &=& \frac{\tilde{S}}{S_0} \label{eq:vczt}
\end{eqnarray}
where the tilde denotes the Fourier transform and $S_0$ is the total source flux. Since the detected image is the convolution of the instrumental point spread function~(PSF) and the source brightness distribution, this leads to:
\begin{eqnarray}
V &=& \frac{\tilde{I}}{S_0\,\tilde{P}} \label{eq:vis}
\end{eqnarray}
where $\tilde{P}$ denotes the Fourier transform of the PSF. In practice, the PSF is measured by observing an unresolved calibrator star, i.e.~a point source, which has unit complex visibility $V=1$.

In this study, we used non-redundant aperture masks, with each baseline pair corresponding to a unique point on the spatial frequency plane. We used masks with 7, 9, and 18 holes, giving 21, 36, and 153 independent baselines, respectively.  Hole diameters and transmission fractions are provided in Table~\ref{table:masks}. The subaperture configurations on the masks were designed to provide a uniform and isotropic sampling of the complex visibility function, with the specific mask chosen to observe a given target depending on the target's brightness and the expected sources of systematic error. For example, although the 18 hole masks had slightly longer baselines than the 7 or 9 hole masks, they had lower total throughput with a broader PSF. This meant that they could only be used with narrow band filters, which restricted their use to brighter targets.

To identify faint companions around our targets, we used a quantity derived from the complex visibility known as the closure phase~$\Theta$ \citep{1958MNRAS.118..276J, 1986Natur.320..595B}. The closure phase is obtained by adding the complex visibility phases around a closure triangle of subapertures. Explicitly, if we denote the measured complex visibility phase between the $i$th and $j$th subapertures as $\varphi_{ij}$, the intrinsic complex visibility phase as $\phi_{ij}$, and a phase error due to atmospheric and instrumental effects across the $i$th aperture as $\eta_{i}$, then we have:
\begin{eqnarray}
 \varphi_{ij} &=& \phi_{ij} + \eta_i  - \eta_j \nonumber \\
 \varphi_{ij} &=& \phi_{jk}+\eta_j - \eta_k \nonumber \\
 \varphi_{ij} &=& \phi_{ki}+\eta_k - \eta_i   
\end{eqnarray}
Importantly, the diameter of the mask holes are chosen to ensure that the wavefront phase variations across each subaperture are approximately constant so that they can be neglected. Combining aperture masking with adaptive optics allows subaperture diameters that are larger than the atmospheric Fried parameter and exposure times that are longer than the atmospheric coherence time to be used, providing a greater through-put of photons. It follows that the $\eta_i$ terms cancel out when we take the closure phase sum:
\begin{eqnarray}
  \Theta_{ijk} &=& \phi_{ij}+\phi_{jk}+\phi_{ki} \label{eq:cp}
\end{eqnarray}
where $\Theta_{ijk}$ is the closure phase of the triangle $ijk$.

The independence of the closure phase quantity from major sources of systematic error allows us to achieve the full interferometric resolution according to the Michelson criterion, which is equal to $\lambda/2B$, where $\lambda$ is the observing wavelength and $B$ is the longest baseline on our mask. This is the smallest angular separation for which two point sources would be fully resolved. Given that the longest baseline of the masks used in this study span nearly the entire telescope aperture, this corresponds to angular scales of roughly half the single-aperture diffraction limit. 

\section{Survey Sample} \label{sec:sample}

\begin{figure}
\epsscale{1.2}
\plotone{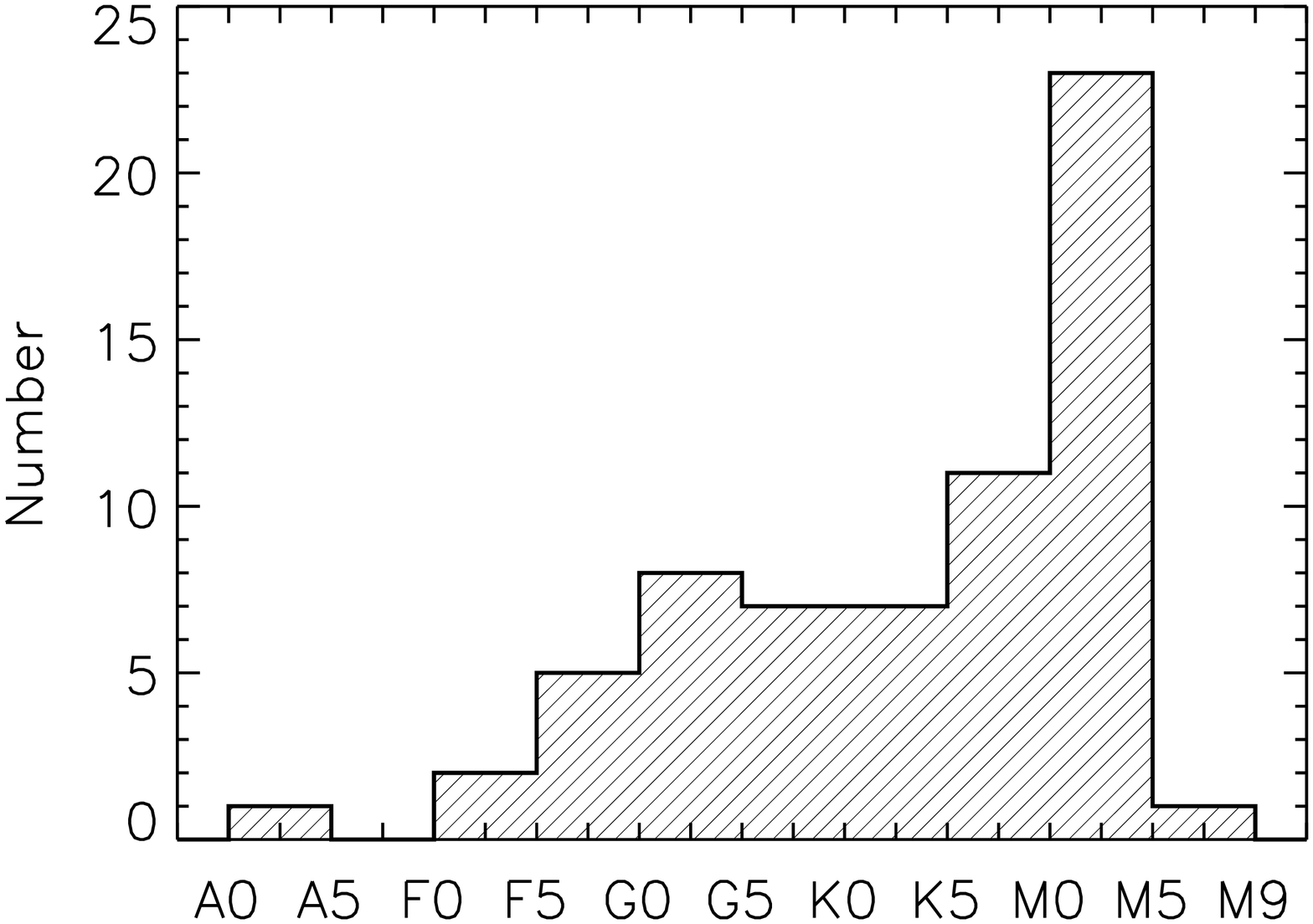}
\plotone{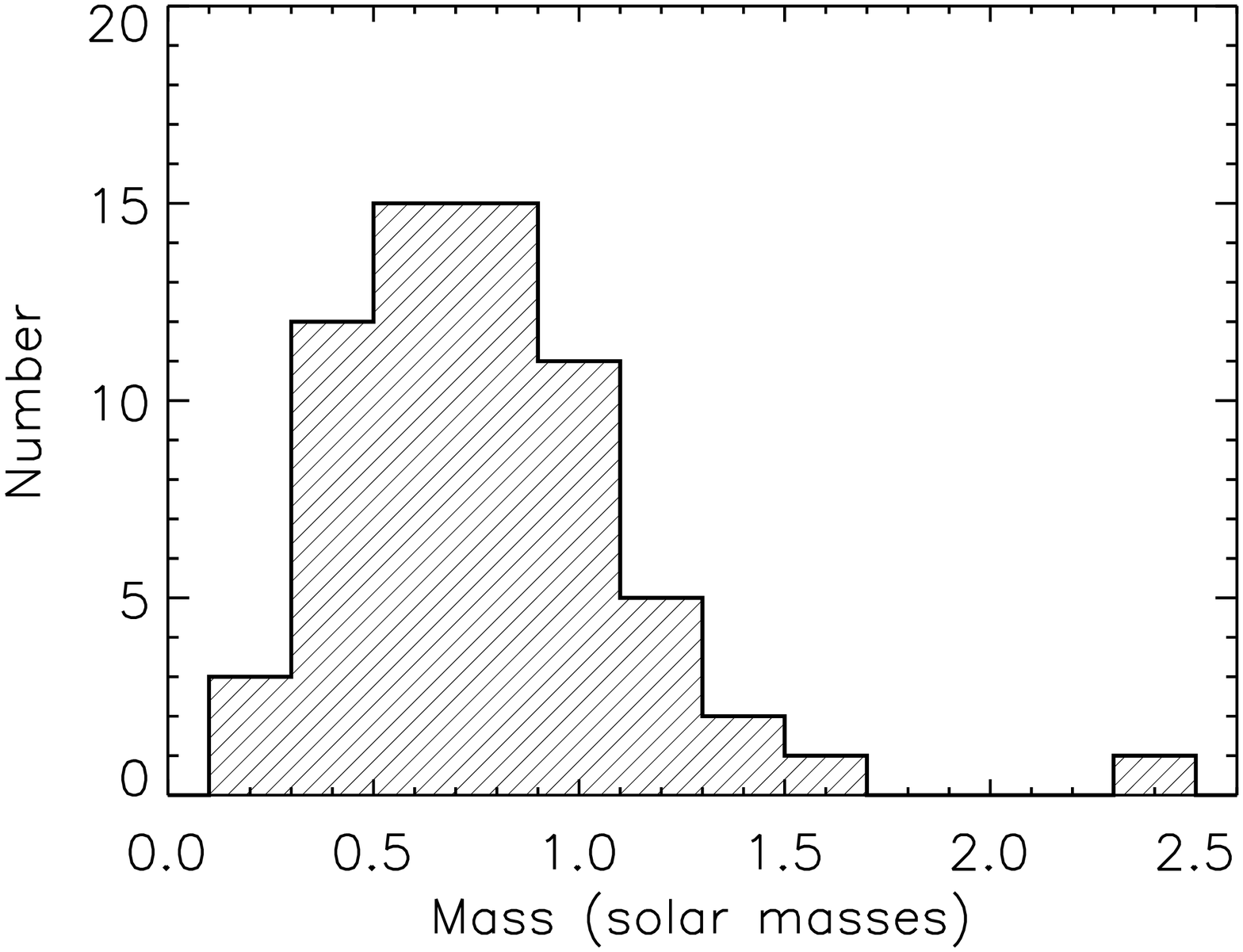}
\caption{The \ssize \ survey targets binned according to the spectral types (top panel) and masses (bottom panel) listed in Table~\ref{table:sample}.    \label{fig:sptmasses}}  
\end{figure}

In 2007, our group initiated a search for close, faint companions around young, nearby stars using the aperture masking facilities installed on the 5.1\,m Hale telescope at Palomar Observatory in California. In subsequent years (2007--2011) the survey was extended and made use of similar facilities installed on the 10\,m Keck~II telescope at Keck Observatory in Hawaii and the 8.2\,m VLT UT4 telescope at the VLT Observatory in Chile.

Our final survey sample consisted of \ssize \ proposed members of the AB Dor \citep{2004ApJ...613L..65Z}, $\beta$~Pic \citep{2001ApJ...562L..87Z}, Her-Lyr \citep{2004AN....325....3F}, Tuc-Hor \citep{2001ApJ...559..388Z}, and TWA \citep{1997Sci...277...67K} moving groups. A concise summary of the sample is provided in Table~\ref{table:mg} while the full list is given in Table~\ref{table:sample}. Figure~\ref{fig:sptmasses} shows the sample members binned according to spectral type and masses. 

In selecting our targets, we noted that many of the moving group members have already been identified as binary systems. The presence of a binary companion within $\sim$\,1\arcsec \ of a target star reduces the ability of aperture masking to detect additional companions, because the interferograms will overlap. Also, similar brightness companions at separations of $\sim$1.5--3\arcsec \ can prevent the adaptive optics system from achieving a stable lock on the target. For these reasons, we chose not to include any targets in our sample that were known to be affected by such issues. 

We also emphasize the difficulty of assigning moving group membership to individual stars. Consequently, it is possible that not all objects in our sample are necessarily young. In particular, the moving group memberships of nine of our targets (HD\,89744, HD\,92945, GJ\,466, EK\,Dra, HIP\,30030, TWA-21, TWA-6, TWA-14, TWA-23) were either unable to be confirmed or else ruled to be unlikely by \cite{2008hsf2.book..757T} using a dynamical convergence analysis.  Furthermore, the existence of Her-Lyr as a genuine moving group is not yet as well-established as the others. To investigate how sensitive our statistical analysis presented in Section~\ref{sec:mc} is to the uncertain membership of these targets, we repeated the calculations separately with them included and then removed from the sample.

\section{Observations and Data Reduction} \label{sec:obsdatared}

We observed our program objects over the course of \nruns \ observing runs using the facility adaptive optics imagers at Palomar (PHARO), Keck (NIRC2), and VLT (CONICA) between April 2007 and April 2011. Each camera has aperture masks installed at (Palomar, VLT) or near (Keck) the pupil stop wheels. The central wavelengths and bandpass widths for each filter used are listed in Table~\ref{table:filters} and details of our observations are summarized in Table~\ref{table:observations}. Observing conditions varied widely, but we attempted to match the observations to the appropriate conditions. Most of our brighter targets were observed through clouds or marginal seeing as they were the only ones we could lock the AO system on, while our fainter targets were typically observed under better conditions. 

Our observing strategy has been described previously in \cite{2008ApJ...679..762K}. Each observation consisted of 1--3 target-calibrator pairs, usually with $\sim$\,10--20 frames per block. We tried to choose calibrators with optical and near-infrared brightnesses that were similar to those of the target, rather than calibrators that were necessarily brighter. This was done due to concerns about the magnitude-dependence of non-common path errors in the adaptive optics system. For targets of brightness $R \lesssim 7$, calibrators were chosen from the stable radial velocity stars of \cite{2002ApJS..141..503N}. For fainter stars, we could not explicitly choose calibrators that had been vetted for close binaries, so we simply chose nearby 2MASS sources with similar colors and brightnesses. In all cases, we tried to select calibrators that appeared to be single and unblended in the 2MASS images, as well as close to the target on the sky ($\lesssim$\,10\,deg for the Nidever et al sources and $\lesssim$\,3\,deg for the 2MASS sources). In addition to reducing overhead times, using nearby calibrators helped to minimize residual wavefront errors introduced by long telescope slews. 

Data reduction was performed using our group's custom-written IDL pipeline (for further details, see \citealt{2006ApJ...650L.131L}, \citealt{2008ApJ...678..463I}, and \citealt{2008ApJ...679..762K}). Complex visibilities were extracted by Fourier-inverting the cleaned data cubes and sampling the $uv$-plane at points corresponding to the mask baselines. Calibration was performed by subtracting the calibrator complex visibility phases from the complex visibility phases of the science targets.

\section{Binary Model Fitting} \label{sec:binaryfitting}

We used the same method as \cite{2008ApJ...679..762K, 2011ApJ...731....8K} to search for companions to our targets over the separation range 20--320\,mas. We only used closure phases in our binary model fitting, discarding the visibility amplitudes as these are more affected by systematic errors. The parameters we fit for were the angular separation $\rho$ between the primary and companion, the position angle $\theta$ of the companion, and the brightness contrast ratio $C=f_p/f_c$, where $f_p$ and $f_c$ were the fluxes of the primary and companion, respectively. Fitting was performed by initially fixing a high contrast ratio of $C=250$ and generating the corresponding model closure phases for each point on a grid of angular separations spanning $20 < \rho < 320$\,mas and position angles spanning $0<\theta<360$\,deg. The point on the $\rho$--$\theta$ grid giving the lowest $\chi^2$ for the measured closure phase values was then taken to be the starting point for a steepest-descent search in which all three model parameters ($C$, $\rho$, $\theta$) were allowed to vary. The initial grid search ensured that the final minimum reached corresponded to the global minimum.

The binary fit was considered to be bona fide if it passed a 99.5\% detection criterion, which has been explained in \cite{2008ApJ...679..762K, 2011ApJ...731....8K}. This was done by generating 10\,000 artificial closure phase data sets with Fourier plane sampling that was identical to that of the measured data. Each artificial closure phase was randomly sampled from a Gaussian distribution with a mean of zero, corresponding to an unresolved point source, and the same variance as the corresponding measured value. A best-fit companion contrast $C$ was then obtained for each set of artificial closure phases using $\chi^2$ minimization at each point on the $\rho$--$\theta$ grid.  Once again, by searching over the entire grid we ensured that the global minimum was identified. A 99.9\% detection threshold was then defined separately for five contiguous annuli (20--40\,mas, 40--80\,mas, 80--160\,mas, 160--240\,mas, 240--320\,mas), corresponding to the 0.1th percentile of the best-fit contrasts obtained for the artificial data sets within that annulus. In other words, if the target was a point source instead of a binary, there was only a 0.1\% chance that the measured closure phases would give a best-fit contrast lower than the threshold value in the annulus corresponding to the best-fit separation. This corresponds to a $5 \times 0.1\%=0.5\%$ false alarm probability across the full 20--320\,mas range. Therefore, if the best-fit model satisfied this condition, the detection was considered to be real at 99.5\% confidence.

It was important to ensure that any high probability ($>$99.5\%) detections were not caused by companions around one of the calibrators rather than around the science target. A small number of such false alarms ($\sim$\,5) did occur during the course of our analysis. Such cases were usually quite straightforward to identify by systematically repeating the calibration and binary fitting, excluding one calibrator at a time. Given that the calibrators did not have known ages, but were likely to be $\sim$\,Gyr old, any companions detected around them were almost certainly not substellar, and so they were not considered further.

\section{Results} \label{sec:results}

Using the method described in Section~\ref{sec:binaryfitting}, we identified stellar companions to four of our AB~Dor targets (HIP\,14807, HD\,16760, HD\,113449, HD\,160934) and report our best-fit binary solutions in Table~\ref{table:binariesdetected}. Of these, the companion to HIP\,14807 is a new discovery while the companions to HD\,16760, HD\,113449, and HD\,160934 are the same as those discovered independently using radial velocity \citep{2009A&A...505..853B, 2009ApJ...703..671S, 2009AIPC.1094..788C, 2010RMxAC..38...34C, 2006Ap&SS.304...59G}. We describe the detected companions in Sections~\ref{sec:hip14807}--\ref{sec:hd160934}, and present our full survey detection limits in Section~\ref{sec:generallimits}.

\subsection{HIP\,14807} \label{sec:hip14807}

A companion was clearly detected in our Keck observations of HIP\,14807 on 2009 November 21 (MJD 55156.2) at an angular separation of $\rho = 28.74 \pm 0.19$\,mas with a contrast ratio of $C = 3.00 \pm 0.06$ in the CO filter. Assuming a system age of $110 \pm 40$\,Myr, interpolation of the NextGen isochrones of \cite{1998A&A...337..403B} gives an estimated companion mass of $0.52 \pm 0.09$\,\ms, which includes the uncertainty in the age and distance, as well as the uncertainty in the fitted contrast.

\begin{figure}
\epsscale{1.2}
\plotone{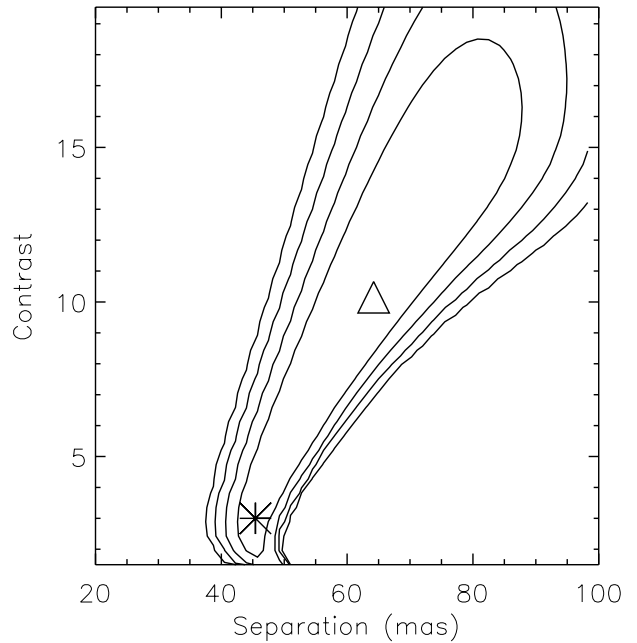}
\caption[]{The contrast/separation degeneracy for the fit to the HIP\,14807 data taken in K band on 2007 Nov 29 (MJD 54433.1), showing the default fit from our pipeline (triangle) and the fit after fixing the contrast (star). The third axis of position angle has been integrated over to give a two-dimensional likelihood plot. Contours are nominally at 90, 99, 99.9, and 99.99\% confidence. \label{fig:hip14807degen}}
\end{figure}

The companion was also detected at high confidence in the Palomar data from 2007 November 29 (MJD 54433.1), with a fitted contrast ratio of $C=10.15 \pm 3.71$. However, this error bar is neither symmetric nor realistic, as there is a strong degeneracy between contrast and separation for small separations in aperture masking data sets. This is illustrated in Figure \ref{fig:hip14807degen} (see also Figure 7 in \citealt{2006ApJ...649..389P}, Table 2 in \citealt{2008ApJ...678..463I}, and Figure 2 in \citealt{2011A&A...528L...7H}). A fuller discussion of this degeneracy is provided in Section 2.1 of \cite{2009ApJ...695.1183M}. In cases such as these, quick data sets were taken with only one or two calibration observations. As a result, the quoted error bars are not necessarily accurate at the few tens of a percent level, because the dispersion between calibrators is used to estimate the errors in the closure phases. Despite this, global orbital fitting to multiple aperture masking data sets has been performed successfully by using a single contrast for all epochs, with the resulting astrometric fits being consistent with those obtained using other techniques, and having reduced $\chi^2$ of order unity (eg.~\citealt{2009ApJ...699..168D}). 

For these reasons, we redid the binary fitting to the MJD 54433.1 data using a prior on the contrast determined from the other well-constrained fit to the MJD 55156.2 data. The results of this revised fit are given in Table~\ref{table:binariesdetected2}.

\subsection{HD\,16760} \label{sec:hd16760}

HD\,16760 is unusual because it shows signs of being both young and old. Its youth is implied by its high lithium abundance, as well as its physical association with the active star HIP\,12635 and a common proper motion with the AB Dor group \citep{2004ApJ...613L..65Z, 2008hsf2.book..757T}. Its old age is implied by its low $v \sin i$ value ($2.8 \pm 1.0$\, km\,s${}^{-1}$, \citealt{2009A&A...505..853B}; $0.5 \pm 0.5$\, km\,s${}^{-1}$, \citealt{2009ApJ...703..671S}) and its low Ca H \& K activity index ($\log R^{\prime}_{HK} =-4.93$, \citealt{2009ApJ...703..671S}; $\log R^{\prime}_{HK} =-5.0 \pm 0.1$, \citealt{2009A&A...505..853B}), which is consistent with field dwarfs ($\log R^{\prime}_{HK}=-4.99\pm 0.07$, \citealt{2008ApJ...687.1264M}) and inconsistent with high probability members of the 625\,Myr Hyades cluster ($\log R^{\prime}_{HK}=-4.47 \pm 0.09$, \citealt{2008ApJ...687.1264M}) and other young stars (see Tables~5--8 of \citealt{2008ApJ...687.1264M}). However, we note that this system is not the only example of a binary pair showing contradictory age indicators: when examining the activity consistency of known binary pairs, \cite{2008ApJ...687.1264M} identified a similar case of an inactive primary with an active companion (HD\,137763 A/B).

Previous radial velocity measurements have shown that HD\,16760 possesses a close companion \citep{2009ApJ...703..671S, 2009A&A...505..853B}, for which Sato and coworkers derived a minimum mass $M_2 \, \sin i$ value of $13.13 \pm 0.56$\,\mj, while Bouchy and coworkers obtained a similar value of $14.3 \pm 0.9$\,\mj. We clearly detected this companion in our Keck data from 2008 December 23 (MJD 54823.2), 2009 August 6 (MJD 55049.6) and 2009 November 20 (MJD 55155.3) (Table~\ref{table:binariesdetected}). Taking the weighted mean of the well-constrained isochrone mass estimates, we obtain a mass of $M_2 = 0.28 \pm 0.04$\,\ms \ for the companion, which includes the uncertainty in the age, distance and fitted contrasts. This places it well within the stellar mass range.

Meanwhile, due to degeneracy between contrast and separation (see Section~\ref{sec:hip14807}), combined with mediocre data quality, the separation derived from the MJD 55155.3 K band data was inconsistent with the separation derived from the J and H band data taken on the same night. For this reason, we obtained a further observation the following night with the CO filter, which has a very similar bandpass to the Kcont filter (see Table~\ref{table:filters}; we had intended to use the Kcont filter, but there was a mix-up in the filter selections). The binary parameters derived from this follow-up observation are in close agreement with the values obtained from the J and H band observations. 

We also repeated the fit to the degenerate K band data with a prior on the contrast obtained by combining the fitted contrasts to the other K band epochs. The system properties derived from this re-analysis are reported in Table~\ref{table:binariesdetected2}, and agree with the values obtained for the J and H band data sets. We note, however, that the calibration error added in quadrature to obtain a reduced $\chi^2$ of unity for this fit was 1.8 degrees. This is unusually large and suggests that the quoted errors for the inferred parameters are likely to be underestimated somewhat.

\begin{figure}
\plotone{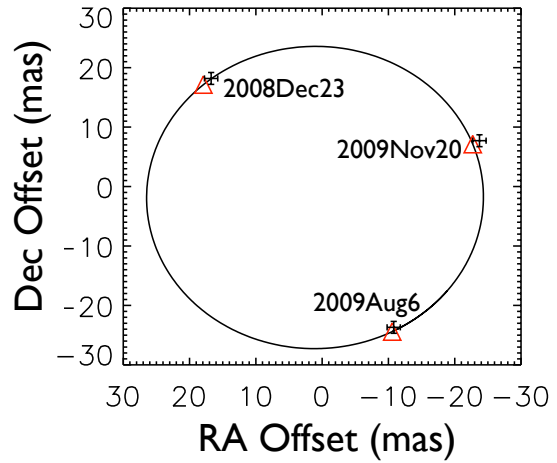}
\caption[HD16760 orbit fit]{Plotted orbital solution for the companion to HD\,16760. The black crosses show the aperture masking astrometry points with associated uncertainties and the red triangles mark the corresponding epochs of the orbital solution. \label{fig:hd16760orbit}}
\end{figure}

Using our multi-epoch data, we were able to derive an orbital solution for the companion. To do this, we fixed the values for the time of periastron $T_0$, orbital period $P$, orbital eccentricity $e$, and argument of periastron $\omega$ published for the radial velocity orbit from \citet{2009ApJ...703..671S}. We were not able to fit for the orbital inclination $i$ using our aperture masking astrometry data because we only measure the axis ratio of the visual orbit, and this varies with the cosine of the inclination. Hence, we are not sensitive to small changes in $i$ when $i$ is near zero, as is the case here. Instead, we combined the model-dependent mass estimate obtained from the aperture masking results with the value for $M_2 \, \sin i$ derived from the radial velocity results to calculate $i= 2.6 \pm 0.5$\,deg. Then with these parameters held fixed, we inferred values for the longitude of the ascending node $\Omega$ and semimajor axis $a$ by fitting to the aperture masking astrometry listed in Tables~\ref{table:binariesdetected} and~\ref{table:binariesdetected2}. The final orbital solution is reported in Table~\ref{table:systemparameters}, and plotted in Figure~\ref{fig:hd16760orbit}. 

Lastly, we note that the rotational velocity of the primary is revised upwards from $v \sin i \sim 0.5$--$4$\, km\,s${}^{-1}$ to $v \sim $20--25\, km\,s${}^{-1}$, a value that is more in line with other members of AB Dor. However, the low Ca H \& K emission of HD\,16760 remains unexplained. The only reason we might expect to see an inclination dependence in the strength of $\log R^{\prime}_{HK}$ is if the Ca H \& K emission varies with latitude on a star, such that polar areas show little emission compared to equatorial regions. We are not aware of any model that would predict this effect.

\subsection{HD\,113449} \label{sec:hd113449} 
 
We detected a companion around HD\,113449 in six of our data sets taken at Palomar and Keck between 2007 Apr 6 and 2010 Apr 25 (Table~\ref{table:binariesdetected}). Four of these data sets (MJD 54196.3, MJD 54634.2, MJD 54821.7, MJD 553311.4) were well-constrained, and taken together imply a companion mass of $0.51 \pm 0.01$\,\ms \ based on the NextGen isochrones of \cite{1998A&A...337..403B}, including the uncertainty in the age, distance and fitted contrasts. The other two data sets, however, gave fits that were degenerate in contrast and separation, as has been described above. For the first of these (MJD 54197.6), we repeated the analysis with a prior on the separation taken from the well-constrained fit to the previous night's data (see footnote in Table~\ref{table:binariesdetected}). For the second case (MJD 54252.1), the analysis was repeated with a prior on the contrast taken from the well-constrained solutions obtained for the four other H band data sets (see Table~\ref{table:binariesdetected2}).

The companion we report here was first announced by \cite{2009AIPC.1094..788C, 2010RMxAC..38...34C} subsequent to the commencement of our survey. Using radial velocity measurements, those authors obtained a value of $F(M_1,M_2,i)=0.0467 \pm 0.0006$\,\ms \ for the spectroscopic mass function and estimated a secondary-to-primary mass ratio of $q=0.57 \pm 0.05$. In addition, using astrometry measurements made with the VLT-I they obtained a value of $i=57\pm 3^\circ$ for the inclination, $\Omega = 124 \pm 4^\circ$ for the longitude of the ascending node and $a = 0.750 \pm 0.030$\,AU for the semimajor axis.

We computed an orbital solution for the companion using our aperture masking astrometry (Tables~\ref{table:binariesdetected} and~\ref{table:binariesdetected2}), allowing $i$, $a$, and $\Omega$ to vary as free parameters in our fitting, while holding $P$, $T_0$, $e$ and $\omega$ fixed at the values determined Cusano and coworkers. However, we found that we could not obtain a reasonable $\chi^2$ value with the period of $P=215.9$ days reported by those authors. Instead, an acceptable fit was made when we  allowed the period to be a free parameter, obtaining $P=216.9$ days. The best-fit parameters are reported in Table~\ref{table:systemparameters} and the corresponding orbit is plotted in Figure~\ref{fig:hd113449orbit}. In particular, our fitted value of $\Omega=202.0 \pm 1.6^\circ$ does not agree with the value of $\Omega = 124 \pm 4^\circ$ reported in \cite{2010RMxAC..38...34C}, but as the details of those VLT-I observations are not given, we cannot make a further comparison. Lastly, the dynamical mass of the system ($M_{\rm{tot}}=1.10 \pm 0.09$\,\ms) appears to be underestimated by $\sim 2\sigma$ when compared to the isochrone-determined masses ($M_1 = 0.84 \pm 0.08$\,\ms \ and $M_2 = 0.51 \pm 0.01$\,\ms). As the orbital period is $\sim$1 year and the astrometric semimajor axis is comparable to the parallax, examining this discrepancy in more detail would require refitting to the raw HIPPARCOS data.

\subsection{HD\,160934} \label{sec:hd160934} 

We detected a companion around HD\,160934 in our Palomar data taken on 2008 June 23 (MJD 54640.3) and Keck data taken on 2010 April 26 (MJD 55312.6) and 2011 April 23 (MJD 55674.6). The binary solutions are all in excellent agreement (Table~\ref{table:binariesdetected}). We obtain a value of $0.54 \pm 0.01$\,\ms \ for the companion mass by combining the estimates from each epoch.

This companion was first reported by \cite{2006Ap&SS.304...59G}, who identified HD\,160934 as a spectroscopic binary with an estimated period of $\sim$\,17.1 years and an eccentricity of $e$\,$\sim$\,0.8. However, these were preliminary values based on limited phase sampling, and a period of approximately half this is also consistent with the data. In fact, this shorter period is the one preferred by \cite{2010Ap&SS.330...47G}, who repeated the fit to the same data with a small number of more modern radial velocity measurements.

\begin{figure}
\epsscale{1.0}
\plotone{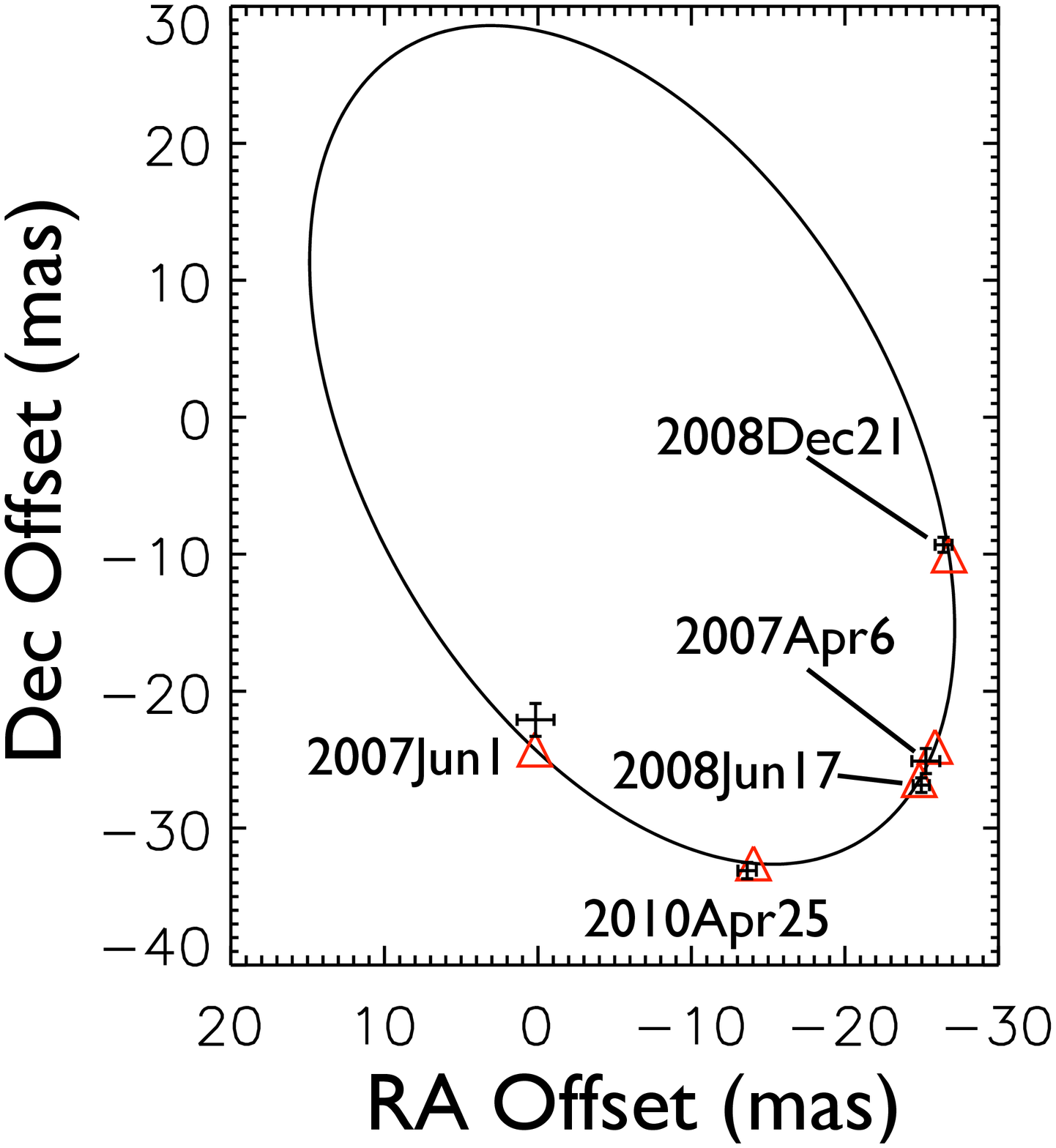}
\caption[HD113449 orbit fit]{The same as Figure~\ref{fig:hd16760orbit}, showing the orbital solution for the companion to HD\,113449. \label{fig:hd113449orbit}}
\vspace{20pt}
\epsscale{1.14}
\plotone{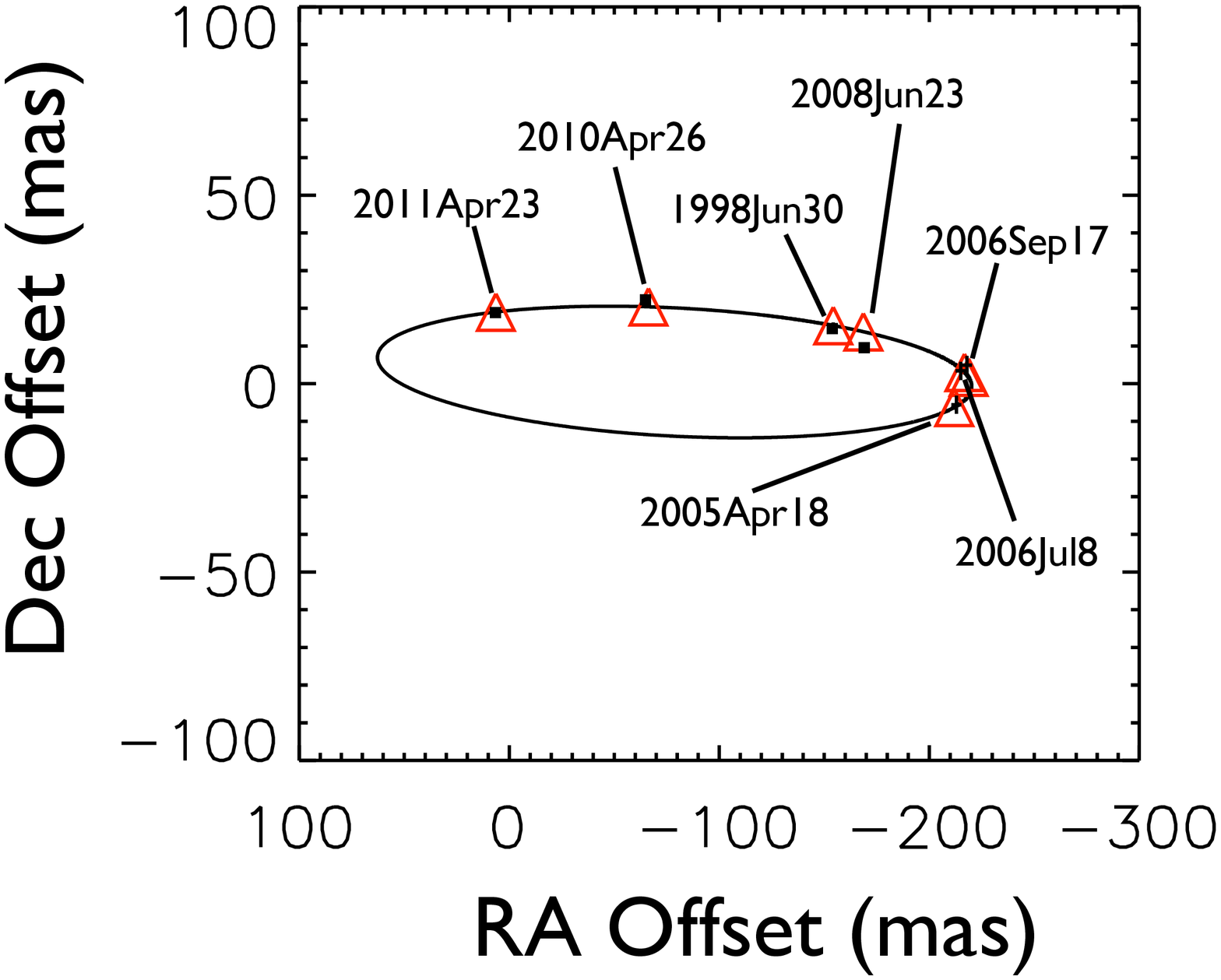}
\caption[HD113449 orbit fit]{The same as Figures~\ref{fig:hd16760orbit} and~\ref{fig:hd113449orbit}, showing the orbital solution for the companion to HD\,160934. Nominal error bars that are too small to be accurately represented are instead contained within filled squares. \label{fig:hd160934orbit}}
\end{figure}

Furthermore, in addition to our values presented in Table~\ref{table:binariesdetected}, relative astrometry measurements have been published by \cite{2007AA...463..707H} and \cite{2007ApJ...670.1367L}. Using the combined data set, which is summarized in Table~\ref{table:hd160934_allastrom}, we performed a least-squares orbital fit and report the results in Table~\ref{table:systemparameters}. The solution is plotted in Figure~\ref{fig:hd160934orbit}. In order to achieve a reduced $\chi^2$ of 1.0, we had to add an extra position angle error of 0.4 degrees to all data, which may indicate a small position angle calibration mismatch between the three instruments used in this fit. Of these parameters, only $T_0$ has an uncertainty that would be significantly changed by the addition of radial velocity data, which have not been made available to us because at least one new paper including those data is in preparation (Montes, private communication). However, we can combine the semiamplitude of the radial velocity curve published in \cite{2010Ap&SS.330...47G} ($K_1$=7.39$\pm$0.22\,km\,s$^{-1}$) with our orbital fit and the HIPPARCOS parallax of 30.2$\pm$2\,mas \citep{2007ASSL..350.....V} to obtain a mass of 0.48$\pm$0.06\,$M_\odot$ for the companion. This value is consistent with the one derived above using isochrones, at the level of the uncertainties.

Although the binary orbit is not taken into account in computing the HIPPARCOS parallax, the period is several times longer than the length of the HIPPARCOS mission and the system was near apastron at the time of observations, so we do not expect the orbital photocenter motion to have a significant effect on the measured parallax. As the parallax uncertainty dominates our mass uncertainty, we have repeated the orbital calculation at several fixed parallax values as given in Table~\ref{table:systemparameters_hd60934_b}. According to the NextGen models of \cite{1998A&A...337..403B}, plausible ages for the companion range from $\sim$50\,Myr through to the zero-age main sequence. Therefore, the dynamical mass does not allow us to place a strong constraint on the system age, but the lower range of allowed values is compatible with the age of AB Dor.

\subsection{Survey Detection Limits} \label{sec:generallimits}

\begin{figure}
\begin{center}
\epsscale{1.25}
\plotone{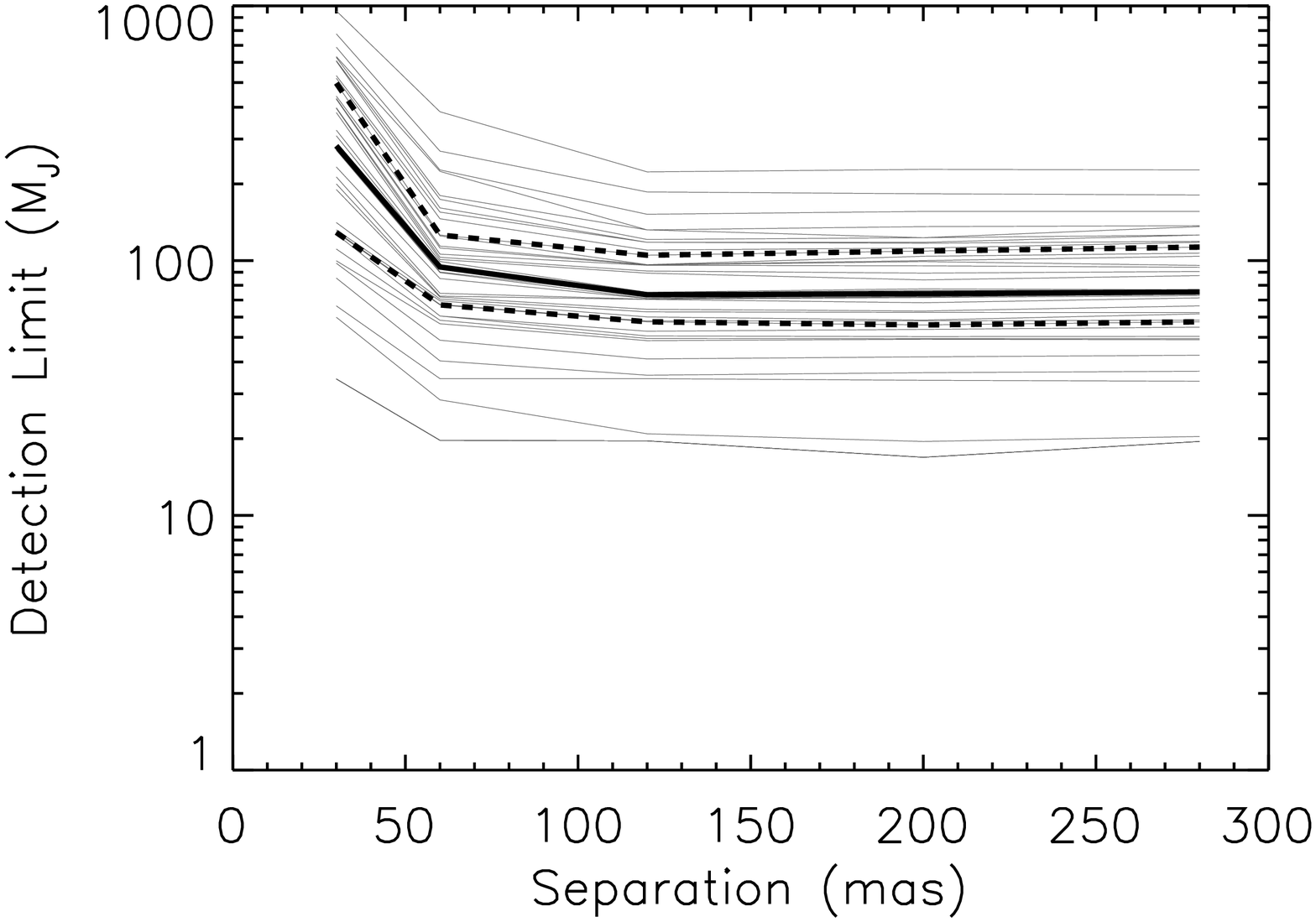}
\plotone{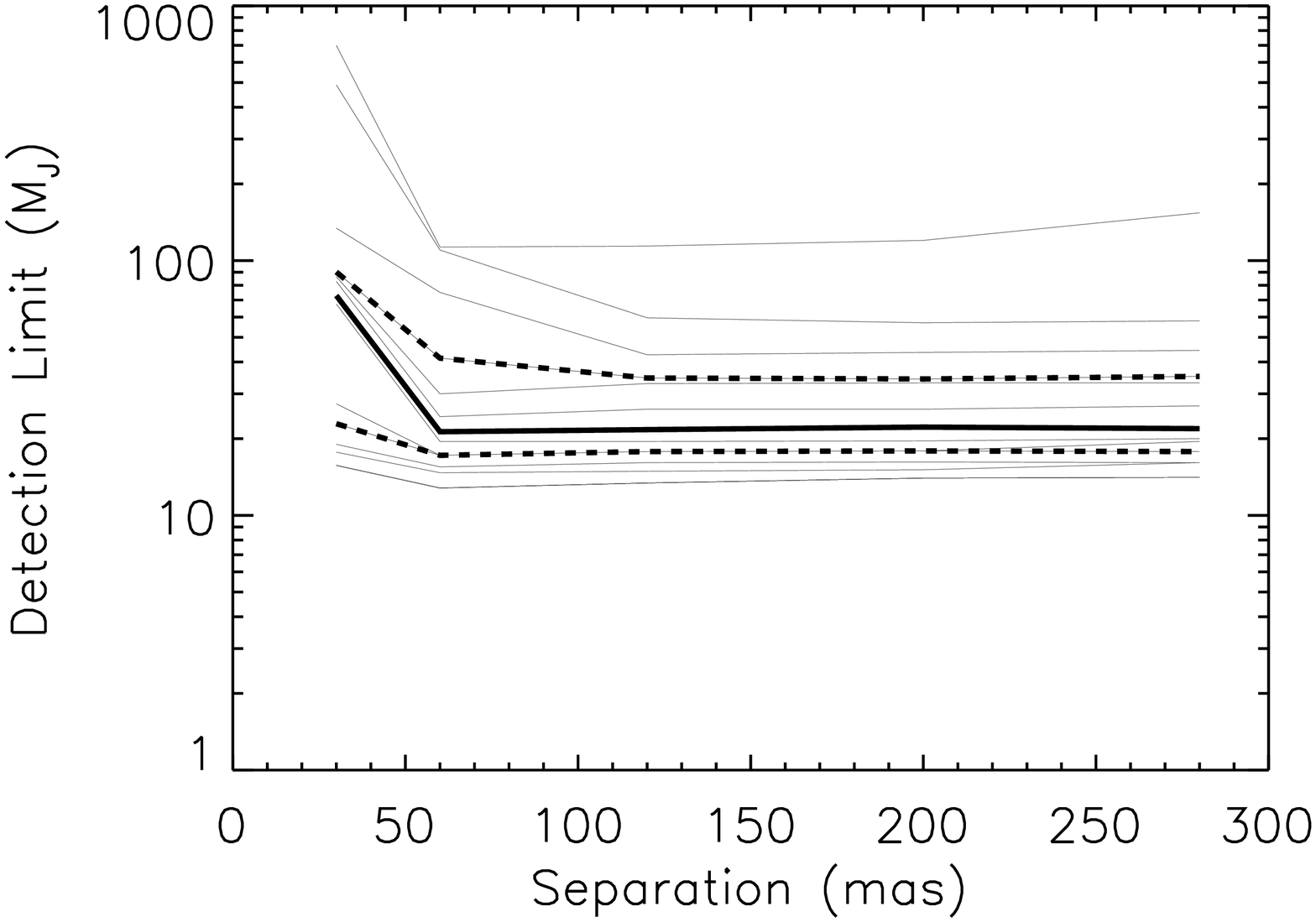}
\plotone{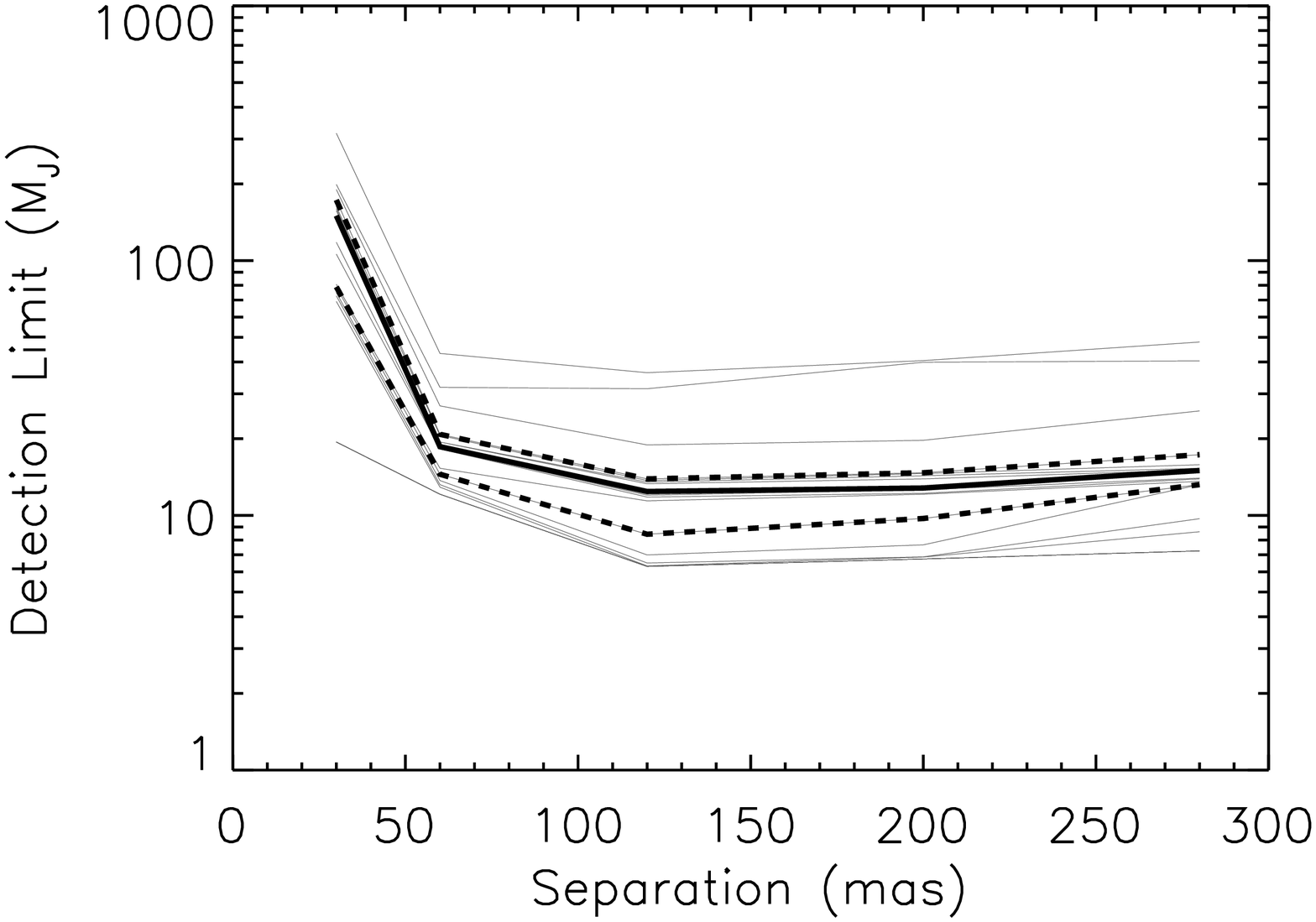}
\caption[Survey detection limits]{Detection limits as a function of angular separation for the AB Dor and Her-Lyr targets (top), the $\beta$~Pic and Tuc-Hor targets (middle), and the TWA targets (bottom). In each plot, solid black lines show the median detection limits within the 20--40, 40--80, 80--160, 160--240, and 240--320\,mas annuli, dashed black lines show the 25th and 75th percentiles, and solid gray lines show the individual detection limits. \label{fig:detlims}}
\end{center}
\end{figure}

We list our detection limits in Table~\ref{table:detlims}, corresponding to the 99.9\% threshold values for each of the  separation annuli, as defined in Section~\ref{sec:binaryfitting}. These were translated into upper limits for companion masses by first converting the contrast ratios into absolute companion magnitudes using the distances listed in Table~\ref{table:sample}. Then combining these intrinsic luminosities with the assumed ages listed in Table \ref{table:mg}, we determined the corresponding companion mass by interpolating an appropriate set of isochrones: specifically, the DUSTY isochrones \citep{2000ApJ...542..464C} for objects with $1400 \, \rm{K} \, \lesssim T_{\rm{eff}} \lesssim 2800 \, \rm{K}$ and the NextGen isochrones \citep{1998A&A...337..403B} for objects with $T_{\rm{eff}} \gtrsim 2800\,\rm{K}$. For the four targets with detected companions (HIP\,14807, HD\,16760, HD\,113449, HD\,160934) we quote the limits obtained for the residual closure phases. 

It should be emphasized that the mass limits quoted in Table~\ref{table:detlims} inherit the systematic errors of the models used to compute them \citep[eg.][]{2002A&A...382..563B}. For instance, \cite{2007ApJ...655..541M} have shown that the predicted luminosities are highly dependent on the treatment of initial conditions, with ``cold start'' models generating luminosities that can be orders of magnitudes fainter than those obtained by the ``hot start'' DUSTY models over Gyr time scales. However, objects in the mass range that our survey is sensitive to would most likely have formed by the gravitational collapse of instabilities in the stellar disk, a process that is more akin to the hot start scenario. Indeed, recent observational evidence appears to favor the hot start models down to masses of $\sim$10\mj \ \citep[eg.][]{2010Sci...329...57L} or even suggest that they could even overpredict the luminosity of such objects \citep{2009ApJ...692..729D, 2010ApJ...721.1725D}. In the latter case, the values quoted in Table~\ref{table:detlims} would be conservative estimates.

With these considerations in mind, Figure~\ref{fig:detlims} shows the detection limits plotted in terms of equivalent companion mass as a function of angular separation. Due to the heterogeneous nature of our observations, which were made using different instruments with different filters, we have divided the targets into three groups in these plots. The top panel shows the detection limits for our older AB Dor ($\sim$110\,Myr) and Her-Lyr ($\sim$200\,Myr) targets, the middle panel shows the detection limits for our younger $\beta$~Pic ($\sim$12\,Myr) and Tuc-Hor ($\sim$30\,Myr) targets, and the bottom panel shows the detection limits for the TWA ($\sim$8\,Myr) targets. The TWA targets have been plotted on their own because all but three of them were observed during the same observing run at VLT using the L${}^\prime$ filter with a 7-hole mask.

\section{Substellar Companion Frequencies} \label{sec:analysis}

We have used our detection limits listed in Table~\ref{table:detlims} to constrain the frequency of $\sim$\,20--80\,$M_{\rm J}$ companions in $\sim$\,3--30\,AU orbits around 0.2--1.5\,\ms \ stars. To do this, we employed the same methodology as \cite{2006AJ....132.1146C}, \cite{2007ApJ...670.1367L}, \cite{2008ApJ...674..466N}, \cite{2009ApJS..181...62M}, and \cite{2010A&A...509A..52C}. We present a brief outline of the approach here, but the 
reader may consult those works for further details.

\subsection{Mathematical Framework}

Firstly, if we denote the outcome of our survey of $N_s$ stars as the set $\{ d_j \}$, where $d_j$ is equal to zero if no companion was detected around the $j$th star or equal to one if a companion was detected, then the likelihood that the fraction of stars with companions is equal to $f$ is given by:
\begin{eqnarray}
P\left( f | \{ d_j \}\right) &=& \frac{\mathcal{L} \left( \{ d_j \} | f \right) \, P(f)}{\int^1_0{\mathcal{L}\left( \{ d_j \} | f \right) \, P(f) \, df}}  \label{eq:bayestheorem}
\end{eqnarray}
where $\mathcal{L} \left( \{ d_j \} | f \right)$ is the likelihood of our data and $P(f)$ is the prior probability that the underlying companion frequency is equal to $f$. We adopt an ignorant prior of $P(f)=1$. 

The fact that we did not detect any 20--80\,$M_{\rm J}$ companions allows us to place an upper limit~$f_u$ on their frequency by integrating Equation~\ref{eq:bayestheorem}, such that:
\begin{eqnarray}
\alpha &=& \frac{ \int^{f_u}_{0}{ \mathcal{L}\left(\{ d_j \} | f \right) \, df} }{ \int^1_0{\mathcal{L}\left(\{ d_j \} | f \right) \, df} }\label{eq:ci2}
\end{eqnarray}
where $\alpha$ is a fraction giving the confidence of our limit (eg.~$\alpha=0.95$ corresponds to a confidence of 95\%). 

Using Poisson statistics, it can be shown that a null result implies:
\begin{eqnarray}
\mathcal{L}\left(\{ d_j \} | f \right) &=& \prod^{N_s}_{j=1}{e^{-f p_j}}
\end{eqnarray}
where $p_j$ is the probability that a substellar companion would have been detected around the $j$th star if there had been one present.

\subsection{Monte Carlo Analysis} \label{sec:mc}

The next task is to determine values for each of the $p_j$ terms, and we did this using a Monte Carlo~(MC) approach. For each target star in our sample, we generated 10\,000 hypothetical companions, each with a mass~$M_2$ and angular separation~$\rho$. The companion masses were either obtained by randomly sampling from an appropriate distribution (see Section~\ref{sec:mdist} below) or else they were set to a fixed value (see Section~\ref{sec:distindependent} below). To obtain the angular separations, we had to properly take into account the companion orbital eccentricities, phases, and orientations. We did this using the approach described by \cite{2006ApJ...652.1572B} in their Appendices 1 and 2. As with the companion masses, this required either randomly sampling these properties from appropriate distributions (see Sections~\ref{sec:edist} and~\ref{sec:adist} below) or else setting them to fixed values (see Section~\ref{sec:distindependent} below).

Having generated 10\,000 hypothetical companions with masses and angular separations for each of the targets in our sample, we then consulted the detection limits in Table~\ref{table:detlims}. Companions with masses that fell above the minimum detectable mass in the corresponding separation annulus for their target star were counted as detections. The $p_j$ value for each target was thus given by the number~$x_j$ of such detections divided by the total number of hypothetical companions generated, i.e.~$p_j=x_j/10\,000$. Equipped with the $p_j$ values, we were then able to calculate an estimate for the companion frequency upper limit~$f_u$  at some level of confidence $\alpha$ by integrating Equation~\ref{eq:ci2}.

\subsection{Mass Distributions} \label{sec:mdist}

The distribution of substellar companion masses in the separation range $\sim$\,3--30\,AU is not yet constrained by observations. To accomodate this uncertainty, we have repeated our MC analysis separately for three different assumed forms for the mass distribution. For the first of these, we extrapolated to 20--80\,$M_{\rm J}$ the power law distribution that has been uncovered by the Keck radial velocity survey for companions with masses $M_2 < 10$\,\mj \ and periods $P$\,$<$\,2000 days \citep{2008PASP..120..531C}, given by:
\begin{eqnarray}
  \frac{d\mathcal{N}}{dM_2} &\propto & M_2^{-1.31}  \label{eq:mpowerlaw}
\end{eqnarray}
where $d\mathcal{N}$ is the number of objects with masses in the interval $[M_2, M_2+dM]$.

The second distribution that we used was the universal mass function proposed by \cite{2009ApJS..181...62M} for companions to solar mass stars, suggested by those authors for companion masses between 0.01\,$M_\odot$ and 1.0\,$M_\odot$ and semimajor axes between 0\,AU and 1590\,AU. It is given by:
\begin{eqnarray}
  \frac{d \mathcal N }{dq} &\propto& q^{-0.39} \label{eq:qpowerlaw}
\end{eqnarray}
where $q$ is the secondary-to-primary mass ratio. 

The last distribution that we used was a log-normal paramaterization proposed by \cite{2008ApJ...679..762K}, derived using an ad hoc physical model of binary formation. It is given by:
\begin{eqnarray}
  \frac{d \mathcal N }{dq} &\propto& \frac{1}{q}\exp{\left[-\frac{1}{2} \left( \frac{\log_{10}q}{\sigma} \right)^2 \right]} \label{eq:qlognormal}
\end{eqnarray}
and we used the authors' proposed value of $\sigma=0.428$.

\begin{figure*}[tH]
\epsscale{1.0}
\plottwo{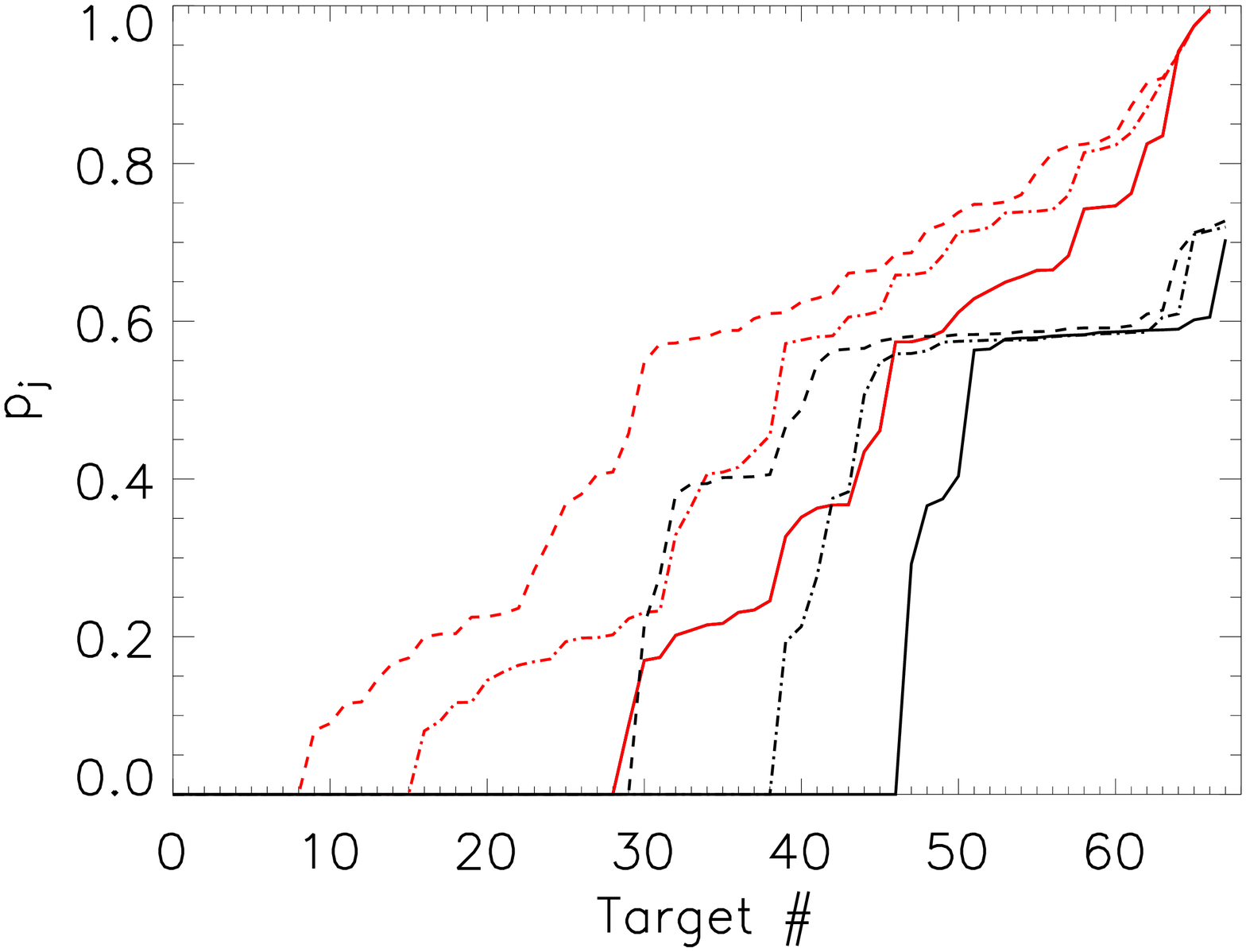}{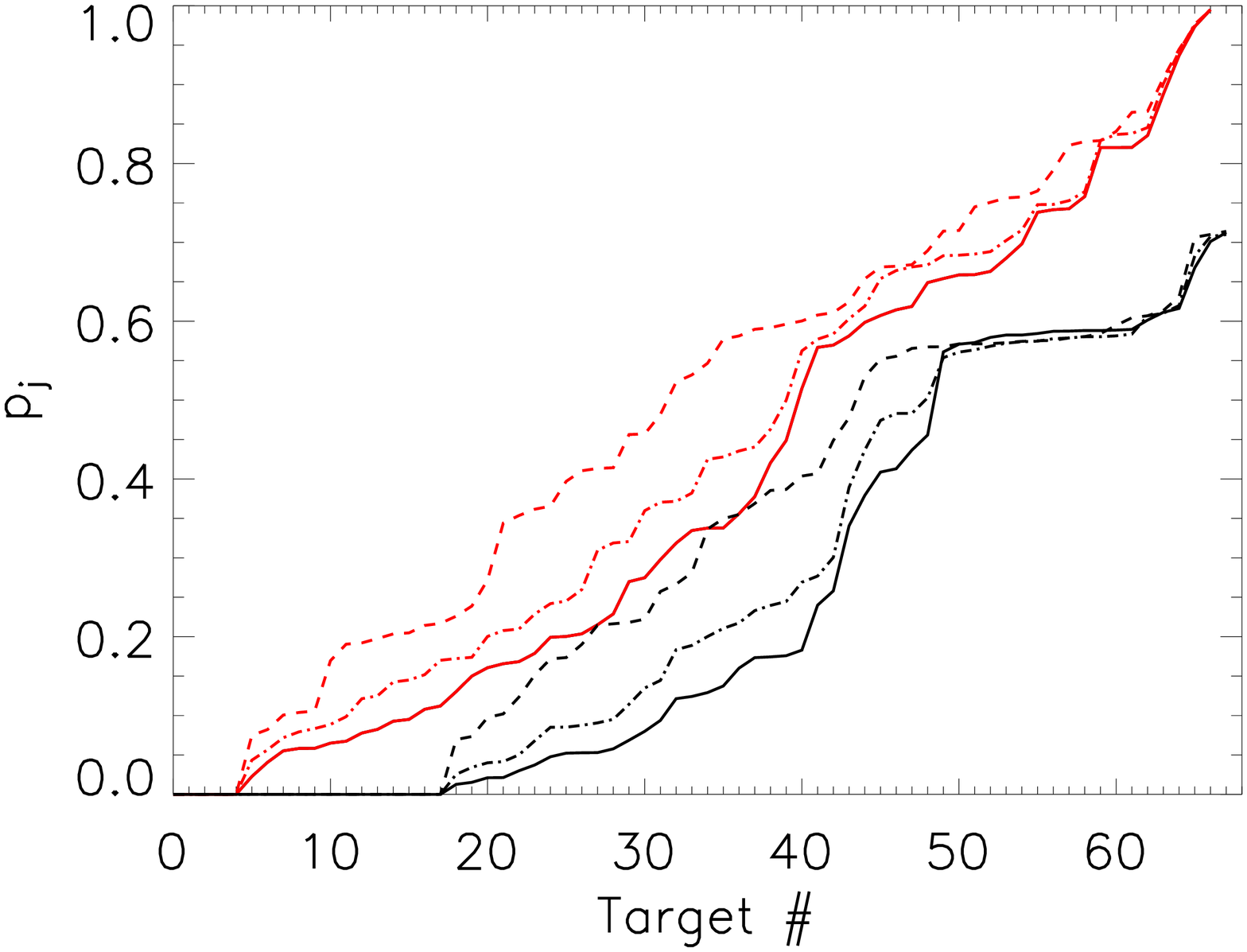}
\caption{Curves showing the $p_j$ values for the case of semimajor axes distributed according to  $d\mathcal{N}/da \propto a^{-1}$ (Equation~\ref{eq:adist}) and eccentricities distributed according to $d\mathcal{N}/de \propto 2e$ (Equation~\ref{eq:edist}). The left panel shows the $p_j$ values obtained for fixed companion masses of 20\,\mj \ (solid lines), 40\,\mj \ (dash-dot lines), and 60\,\mj \ (dashed lines), while the right panel shows the same values obtained for three assumed companion mass distributions: a mass power law given by Equation~\ref{eq:mpowerlaw} (solid lines), a mass ratio power law given by Equation~\ref{eq:qpowerlaw} (dash-dot lines), and a log-normal mass ratio parameterization given by Equation~\ref{eq:qlognormal} (dashed lines). In both panels, the black curves indicate the values obtained using the aperture masking limits only, while the red curves show the equivalent values obtained when the imaging limits are also included (see Section~\ref{sec:previmaging}). Note that for each curve, the values are arranged in ascending order so that, in general, a point on the horizontal axis does not correspond to the same target for all curves. \label{fig:pjvals}}
\end{figure*}

\subsection{Eccentricity Distributions} \label{sec:edist}

As with masses, the distribution of substellar companion orbital eccentricities in the semimajor axis range $\sim$\,3--30\,AU is not yet constrained by observations. We chose to draw companion eccentricities from a distribution of the form:
\begin{eqnarray}
  \frac{d \mathcal N }{de} &\propto& 2e \label{eq:edist}
\end{eqnarray}  
which, as noted in Appendix 2 of \cite{2006ApJ...652.1572B}, is an approximation that has been derived from physical considerations. 

However, to test how sensitive our results were to the distribution of companion eccentricities, we repeated all of our MC analyses for two limiting cases: (1)~fixing the orbital eccentricity of all hypothetical companions to $e=0.9$; (2)~fixing all hypothetical companion eccentricities to $e=0$. 

\subsection{Semimajor Axis Distributions} \label{sec:adist}

We drew substellar companion semimajor axes from an inverse power law of the form:
\begin{eqnarray}
  \frac{d \mathcal N }{da} &\propto& a^{-1} \label{eq:adist}
\end{eqnarray}
over the separation range 3--30\,AU. This distribution was also used by \cite{2009ApJS..181...62M} for $a>30$\,AU (see their Appendix 2 for a discussion) and it is consistent with recent results for stellar binaries between $\sim 5$--$500$\,AU \citep[eg.][]{2008ApJ...679..762K, 2011ApJ...731....8K}. Furthermore, in the event that $>$10\,$M_{\rm J}$ objects can form by the same mechanism as lower-mass gas giant planets, Equation~\ref{eq:adist} is a reasonable extrapolation from the results of \cite{2008PASP..120..531C}, who found a nearly-log-flat distribution for $<$10\,$M_{\rm J}$ gas giant planets at separations $a<3$\,AU.

\subsection{Distribution-independent Approach} \label{sec:distindependent}

In addition to assuming specific forms for the distribution of companion properties, we repeated the analysis with them set to fixed values. This allowed us to obtain conservative upper limit estimates for the companion frequencies. For instance, the less massive a companion is, the more difficult it is to detect because it is fainter and hence the required contrasts are higher. Therefore, by setting all of our hypothetical substellar companions to have some mass $M^\prime_2$, the subsequent result we obtain from the MC analysis is an upper limit on the frequency of all companions with masses $M_2 \geq M^\prime_2$.

Similarly, our ability to detect companions varied with angular separation (Figure~\ref{fig:detlims}), which is related to the semimajor axis of the companion via the distance to the system and the orientation of the orbit. Now suppose we fix the semimajor axes of the hypothetical companions to a certain value~$a^\prime$ and repeat the MC analysis for values over some interval $a^\prime$\,$\in$\,$[a_1,a_2]$. Then the maximum value of $f_u$ obtained from these analyses is the most conservative upper limit estimate for the frequency of companions with semimajor axes on that interval. 

We present the results of these distribution-independent calculations in Section~\ref{sec:mcresults}, as well as the results obtained by assuming the specific distribution forms described in Sections~\ref{sec:mdist}--\ref{sec:adist}.

\subsection{Previous Imaging Observations} \label{sec:previmaging}

Ideally, when performing the calculations described above, we would like to combine the results of our aperture masking survey with those of other imaging surveys targeting wider angular scales. This would allow us to put tighter constraints on the companion frequencies across a larger range of separations. To this end, we identified 49 of our targets that have previously been observed as part of published direct imaging surveys and list these in Table~\ref{table:imassumpt}. For each of these targets, we quote the inner separation angle that was probed by the imaging observations as well as the corresponding sensitivity of the observations. In most cases, these values were taken directly from the published survey limits, but when these were not provided explicitly we attempted to make conservative estimates. We also list each of the sensitivities in Table~\ref{table:imassumpt} as an equivalent minimum detectable companion mass, calculated by interpolating the DUSTY isochrones of \cite{2000ApJ...542..464C} in the same manner outlined in Section~\ref{sec:generallimits}. We incorporated these limits into our analysis described in Section~\ref{sec:mc} by treating hypothetical companions as ``detected'' (i.e.~by increasing $x_j$ by 1) whenever they came within the detectability range of the previous imaging observations (i.e.~if they had separation and masses above the values quoted in Table~\ref{table:imassumpt}). In the next sections, we present the results obtained from this combined approach (aperture masking + previous imaging) together with the results obtained using the aperture masking limits alone. 

\subsection{Calculated $p_j$ Values} \label{sec:pjvalues}

The $p_j$ values calculated separately for each of the three companion mass distributions that we considered  (Equations~\ref{eq:mpowerlaw}, \ref{eq:qpowerlaw}, and \ref{eq:qlognormal}) are plotted in ascending order in Figure~\ref{fig:pjvals} for the case of a companion orbital eccentricity distribution given by $d \mathcal{N} /de \propto 2e$ (Equation~\ref{eq:edist}) and semimajor axis distribution given by $d \mathcal{N} / da \propto a^{-1}$ (Equation~\ref{eq:adist}). In this figure, we immediately see the advantage of combining our aperture masking results with the results from imaging surveys: the overall effect is roughly equivalent to an upwards shift of the $p_j$ values by $\sim$\,10--30\% for the majority of targets.

\subsection{Calculated $f_u$ Values} \label{sec:mcresults}

\begin{figure*}[tH]
\epsscale{1.17}
\plottwo{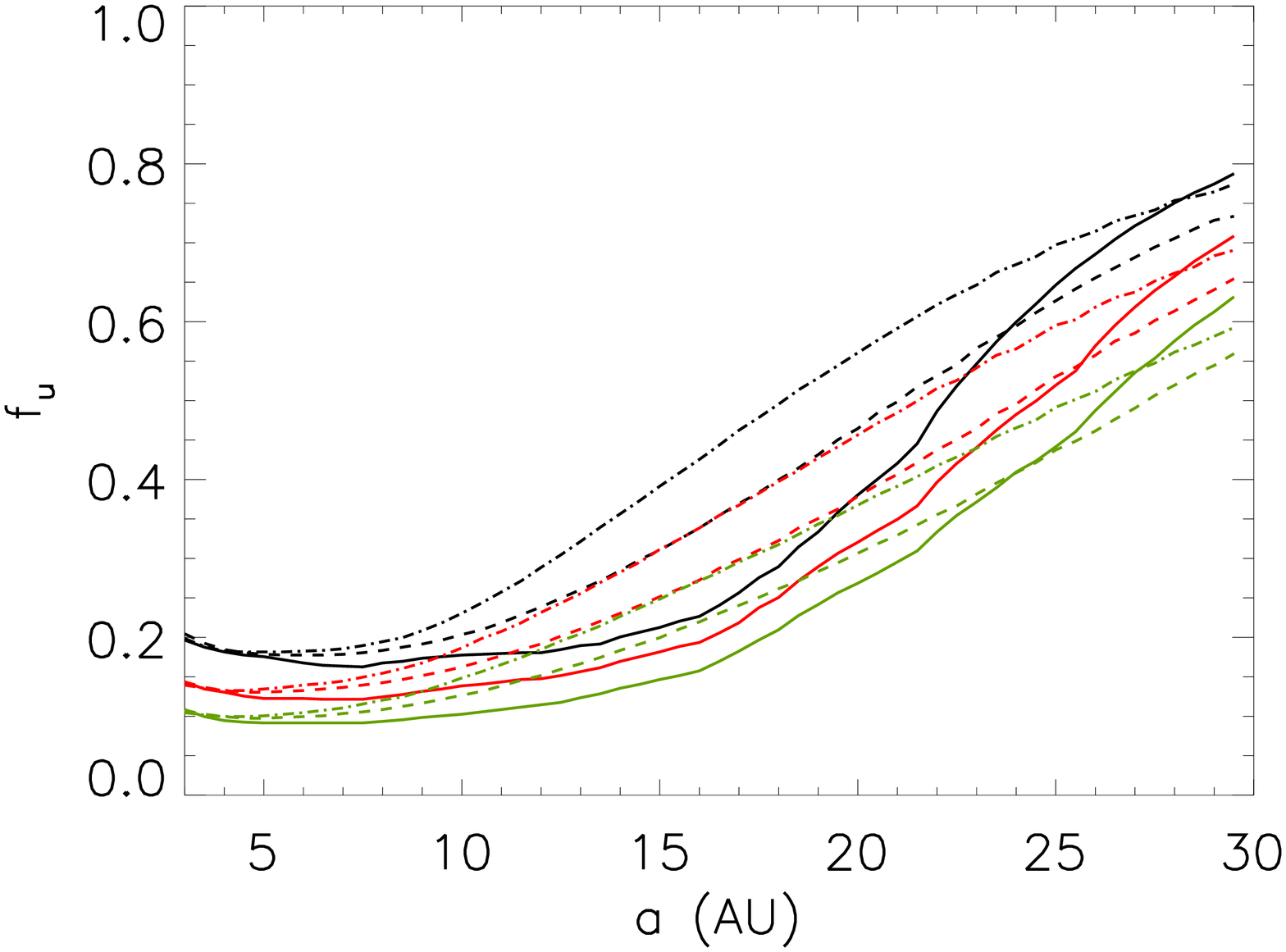}{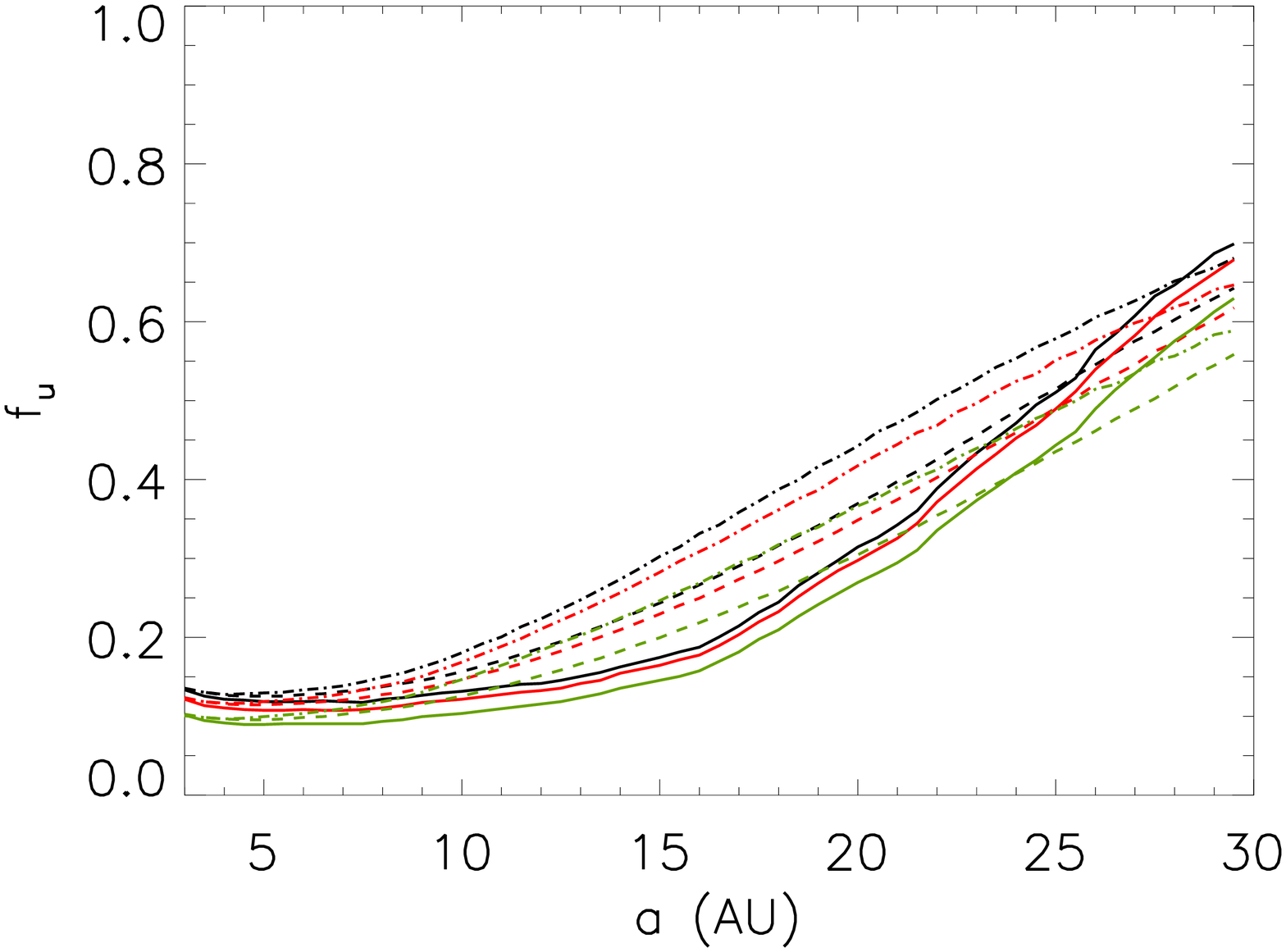}
\caption{95\% confidence $f_u$ estimates as a function of fixed companion semimajor axis obtained for the aperture masking limits alone with different assumptions about the companion eccentricities. The left panel shows values obtained for fixed companions masses of 20\,\mj \ (black lines), 40\,\mj \ (red lines), and 60\,\mj \ (green lines). The right panel shows the same values obtained for companions distributed according to the mass power law given by Equation~\ref{eq:mpowerlaw} (black lines), the mass ratio power law given by Equation~\ref{eq:qpowerlaw} (red lines), and the log-normal mass ratio parameterization given by Equation~\ref{eq:qlognormal} (green lines). In both panels, $f_u$ values are shown for the following cases: fixed companion eccentricities of $e=0$ (solid lines); fixed companion eccentricities of $e=0.9$ (dash-dot lines); and companion eccentricities randomly sampled from a distribution of the form $d\mathcal{N}/de \propto 2e$ (Equation~\ref{eq:edist}) (dashed lines). \label{fig:mcecc}}
\end{figure*}

\subsubsection{Assuming $d\mathcal{N}/da \propto a^{-1}$} \label{sec:mcresults1}

In Table~\ref{table:mcresults}, we present 95\% confidence (i.e.~$\alpha=0.95$ in Equation~\ref{eq:ci2}) upper limit estimates $f_u$ for the frequency of 20--80\,$M_{\rm J}$ substellar companions in the separation range 3--30\,AU, with companion semimajor axes randomly drawn from the inverse power law distribution $d \mathcal{N} / da \propto a^{-1}$ (Equation~\ref{eq:adist}). Also presented are calculations made separately for each permutation of the companion mass and eccentricity distributions described in Sections~\ref{sec:mdist} and~\ref{sec:edist}, respectively, as well as for different fixed companion masses of 20\,$M_{\rm J}$, 40\,$M_{\rm J}$, and 60\,$M_{\rm J}$ (see Section~\ref{sec:distindependent}).

All calculations reported in Table~\ref{table:mcresults} are reasonably robust to the different assumptions made for the companion eccentricities, with the calculated upper limits only differing by $\lesssim$1--2\% depending on whether all companion eccentricities are fixed to $e=0$ or $e=0.9$, or if they are drawn randomly from a distribution of the form $d \mathcal{N} /de \propto 2e$ (Equation~\ref{eq:edist}). When the previous imaging observations are incorporated into the calculations and the ages and distances listed in Tables~\ref{table:mg} and \ref{table:sample} are used, the upper limit estimates vary between 9--12\%, depending on which form is assumed for the distribution of companion masses, but irrespective of what is assumed about the orbital eccentricities. When the previous imaging observations are not included in the analysis, the equivalent limits vary between 13--19\%. For fixed companion masses of 20\,\mj, 40\,\mj \ and 60\,\mj, the upper limit estimates vary between 15--16\%, 11--12\% and 9--10\%, respectively, when the imaging observations are included, and between 25--27\%, 18--20\% and 14--15\%, respectively, when the imaging observations are not included.

To investigate how sensitive our results are to the uncertainties in the distances and ages of our targets, we repeated the above calculations using the $1\sigma$ upper limits for the ages and distances of each target provided in Tables~\ref{table:mg} and \ref{table:sample}. For instance, instead of using a distance of 28\,pc and an age of 110\,Myr for PW~And, we used $28+7=35$\,pc and $110+40=150$\,Myr, respectively. Assuming upper values for the ages and distances in this way results in a downward revision of our sensitivities to faint companions at smaller separations. Therefore, we had to re-calculate the survey detection limits presented in Table~\ref{table:detlims} before repeating the analysis described in Sections~\ref{sec:mc}--\ref{sec:previmaging}. Depending on which of the companion mass distributions is used, the upper limit estimates obtained from this analysis vary between 11--15\% when the imaging observations are included, and between 17--24\% when the imaging observations are not included. For fixed companion masses of 20\,\mj, 40\,\mj \ and 60\,\mj, when the imaging observations are included the calculated upper limits vary between 20--23\%, 14--15\% and 11--12\%, respectively, and when the imaging observations are not included they vary between 38--40\%, 24--25\% and 18--20\%, respectively.

We also investigated how sensitive our results are to the 9 targets of less certain membership identified in Section~\ref{sec:sample} (i.e.~HD\,89744, HD\,92945, GJ\,466, EK\,Dra, HIP\,30030, TWA-21, TWA-6, TWA-14, TWA-23) by removing them and the 7 Her-Lyr targets from the analysis. When all 16 of these targets are removed and we randomly sample the companion masses, the upper limit estimates vary between 12--15\% when the imaging observations are included and between 16--23\% when the imaging observations are not included, depending on which of the three companion mass distributions from Section~\ref{sec:mdist} is used. For fixed companion masses of 20\,\mj, 40\,\mj \ and 60\,\mj, the upper limit estimates vary between 18--20\%, 14--15\% and 12--13\%, respectively, when the imaging observations are included, and between 31--32\%, 22--23\% and 17--19\%, respectively, when the imaging observations are not included. 

\begin{figure*}[H]
\epsscale{1.15}
\plottwo{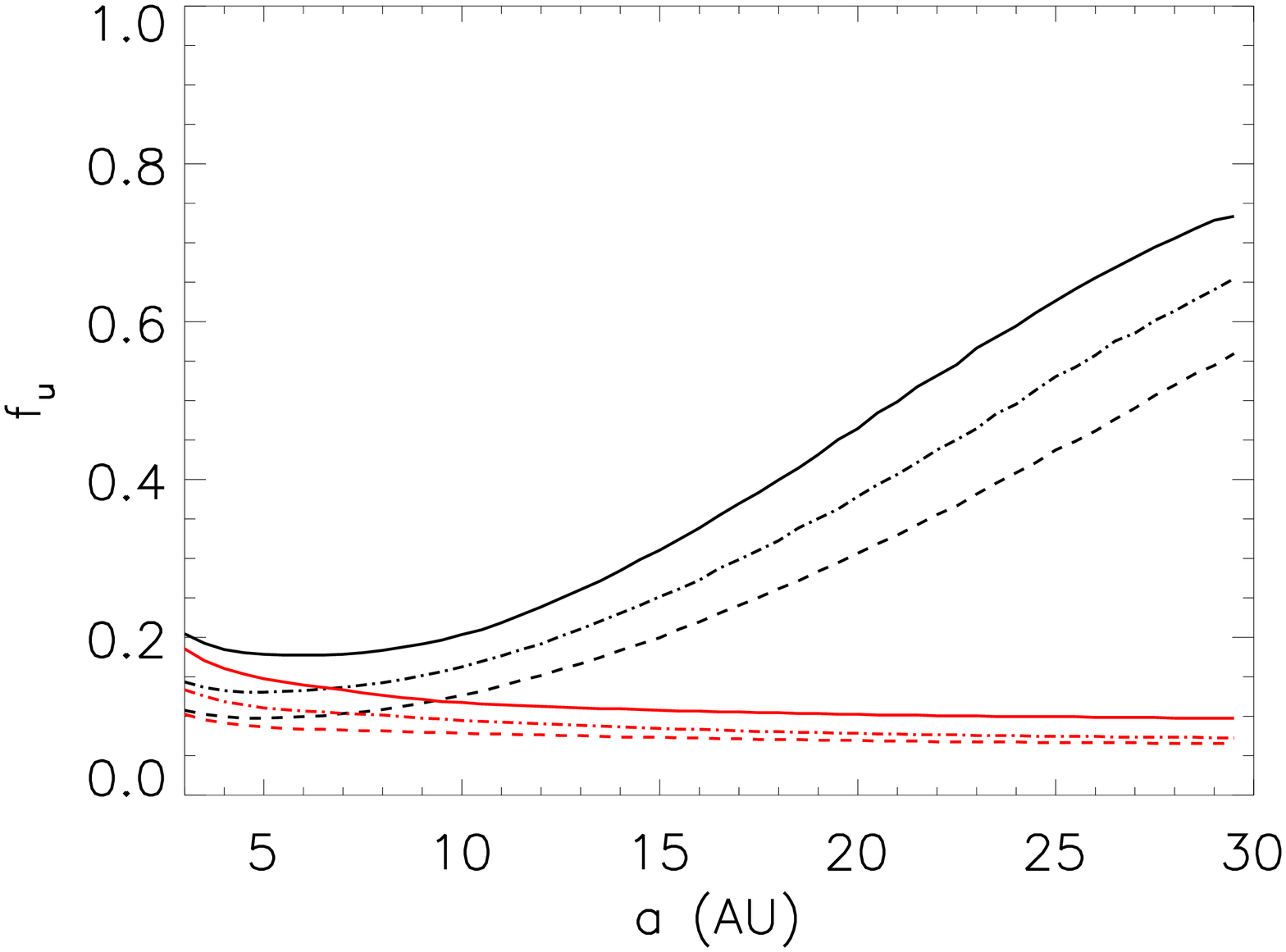}{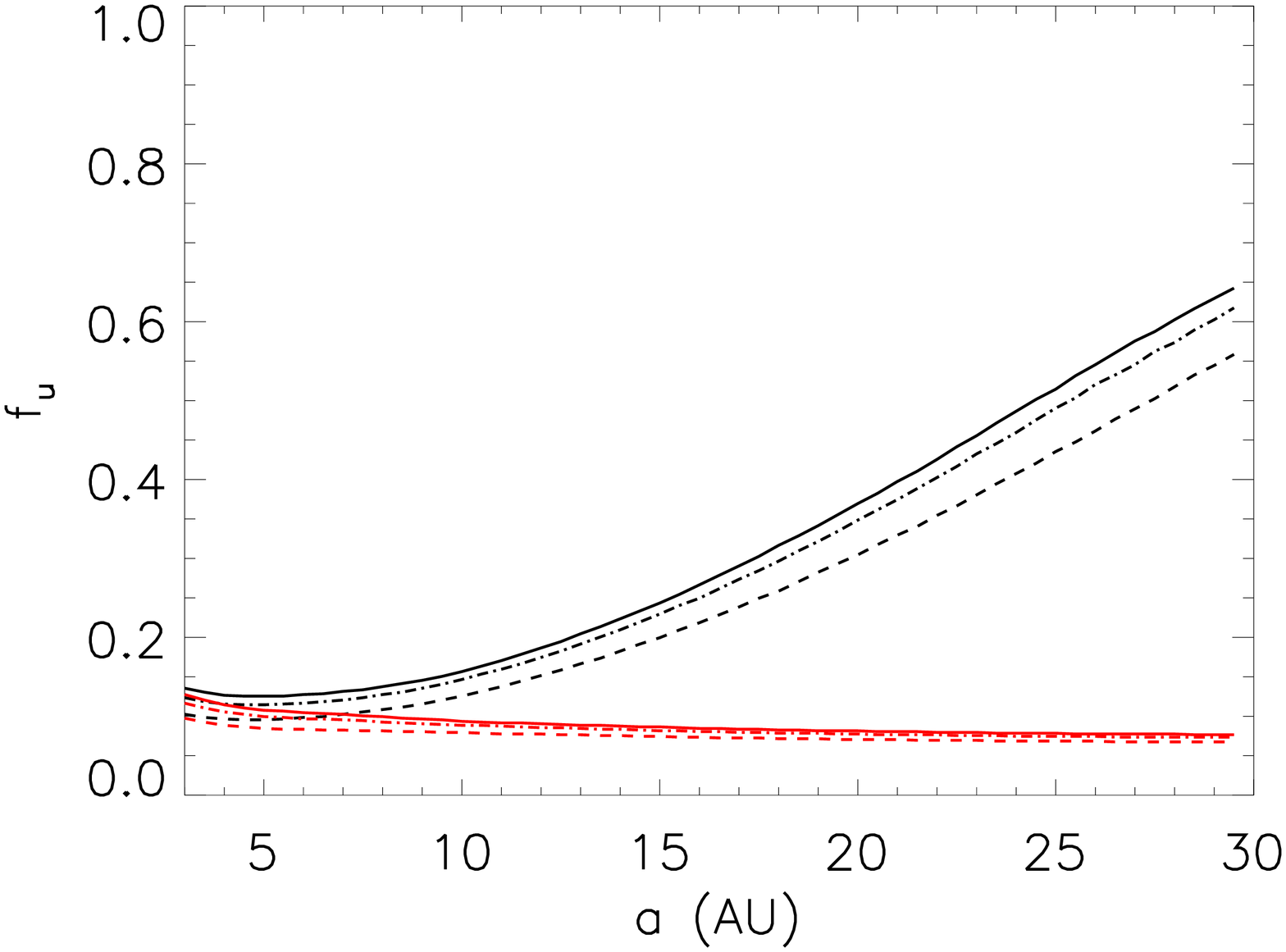}
\caption{95\% confidence $f_u$ estimates for companions with masses in the range 20--80\,$M_{\rm J}$ as a function of fixed companion semimajor axes. As in Figure \ref{fig:pjvals}, the left panel shows the results for fixed companion masses and the right panel shows the results for companion masses drawn randomly from distributions. Colors and linestyles are the same as Figure~\ref{fig:pjvals}. \label{fig:mc1}}
\epsscale{1.15}
\plottwo{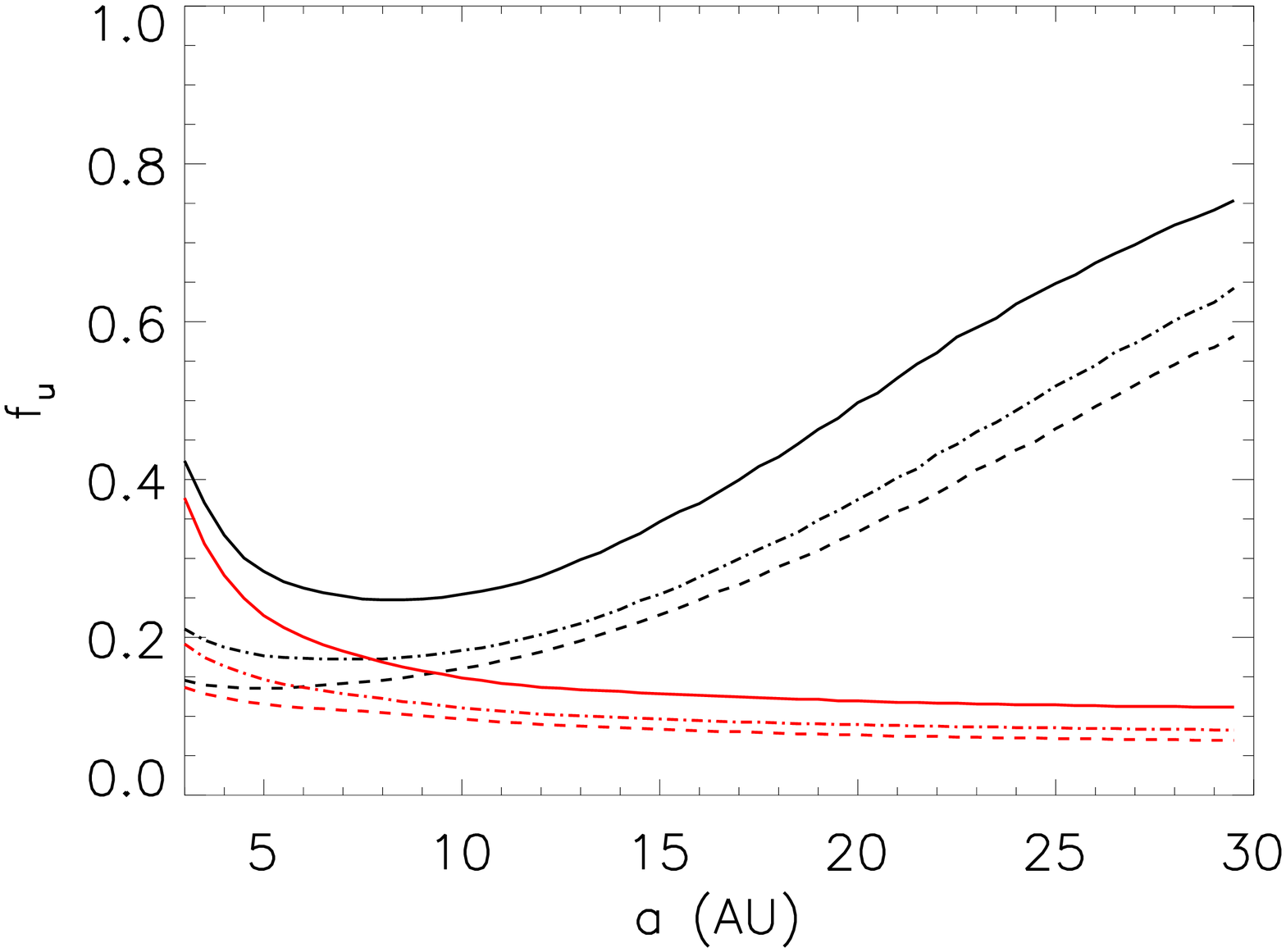}{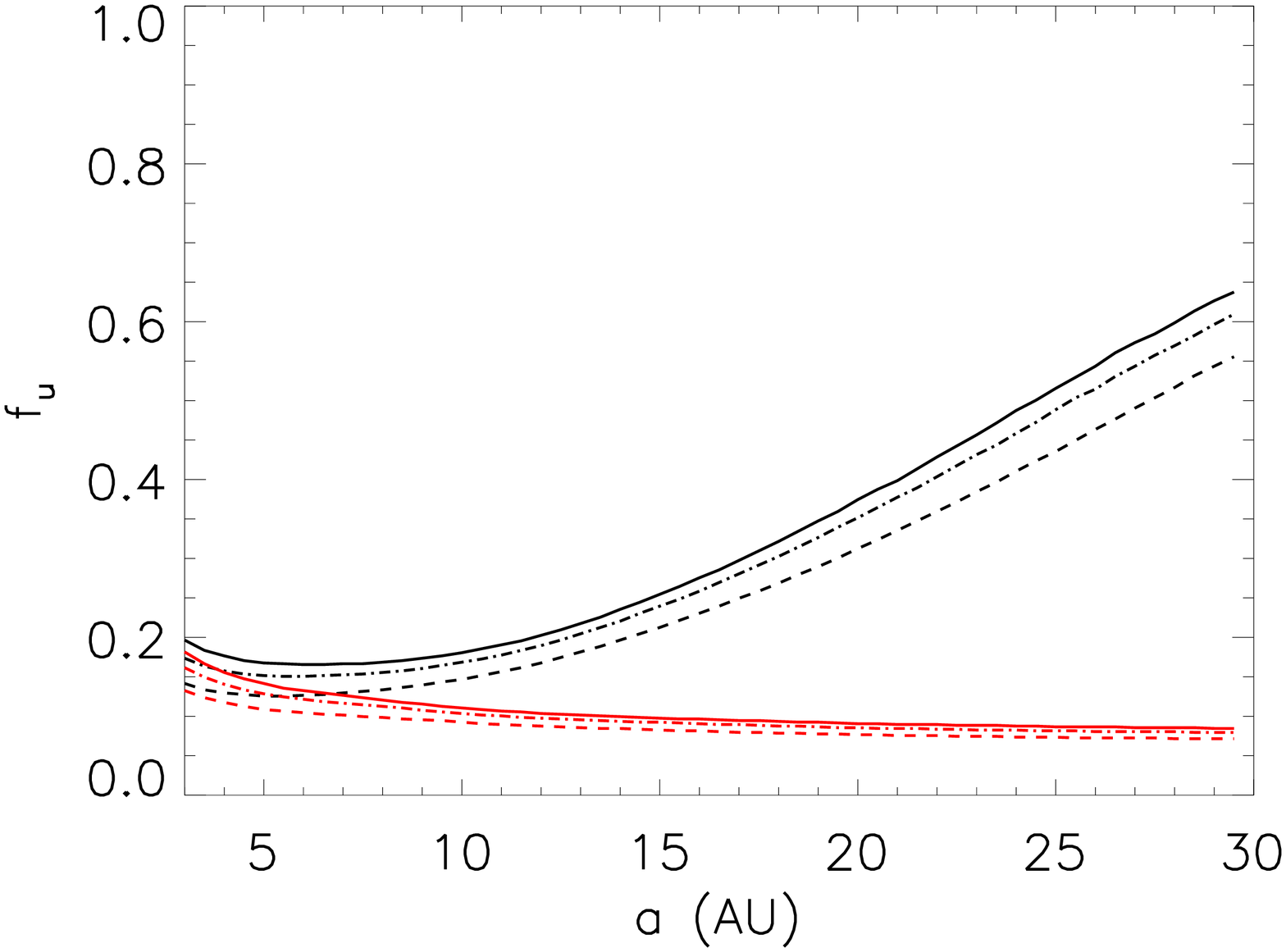}
\caption{ The same as Figure~\ref{fig:mc1}, but using values for the target distances and ages that give conservative estimates for the companion detection sensitivities. Specifically, the upper $1\sigma$ limits given in Tables~\ref{table:mg} and~\ref{table:sample} were used for the target ages and distances, respectively.  \label{fig:mc2}}
\plottwo{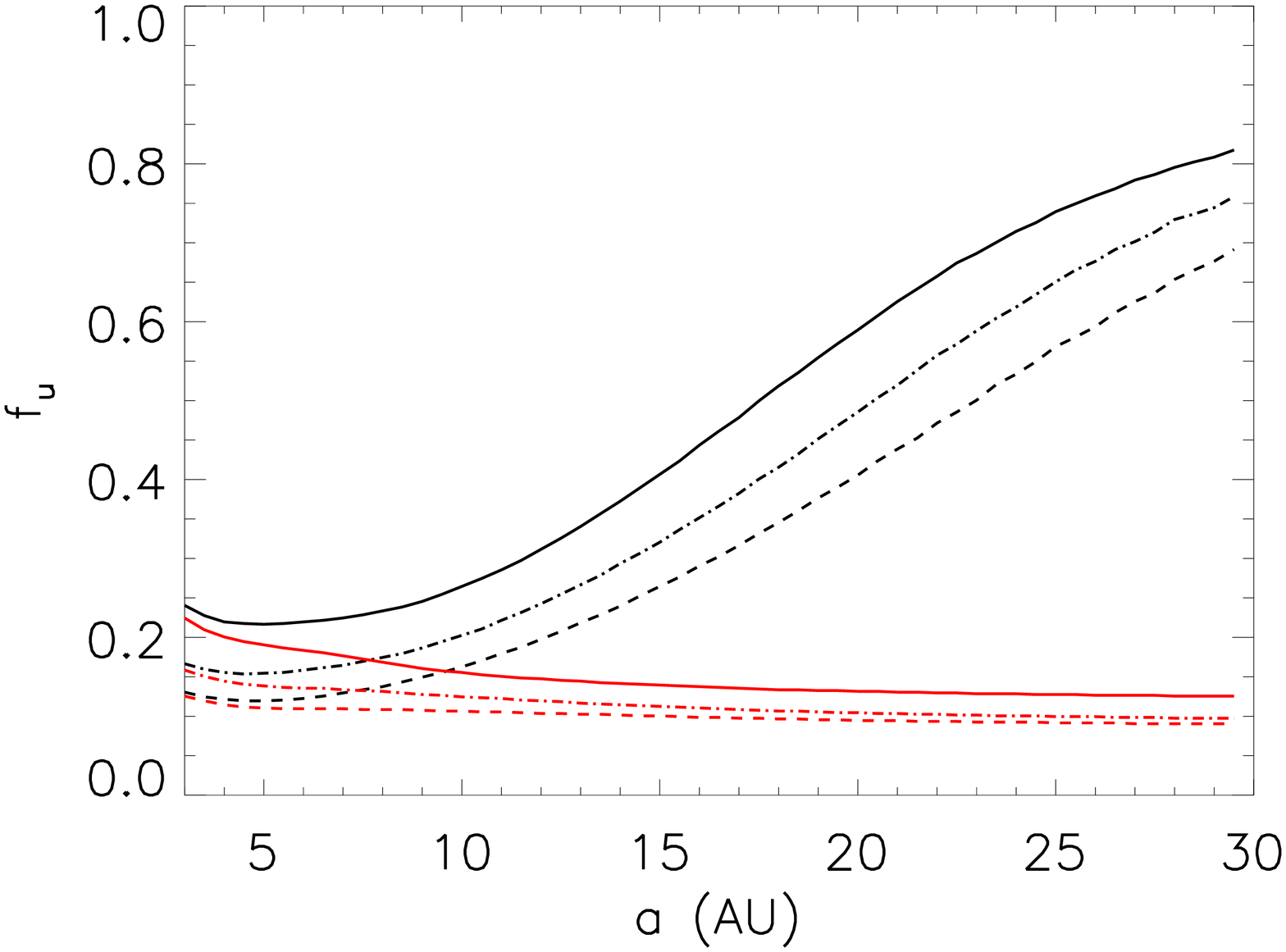}{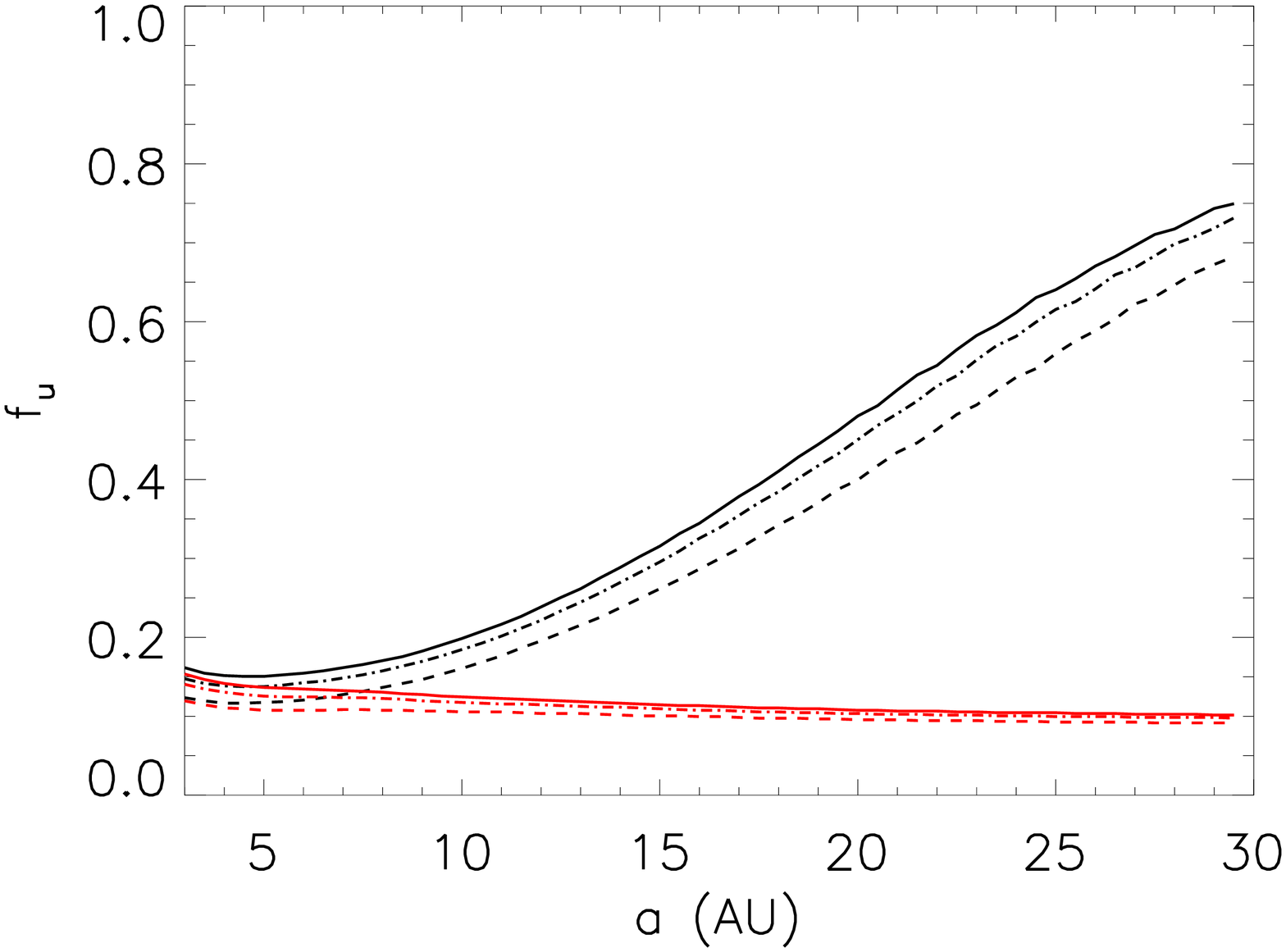}
\caption{The same as Figure~\ref{fig:mc1}, but excluding 16 of the targets whose moving group membership is not well-established; namely, the 9 targets that did not meet the `high probability' criterion of \cite{2008hsf2.book..757T} in their dynamical convergence analysis (HD\,89744, HD\,92945, GJ\,466, EK\,Dra, HIP\,30030, TWA-21, TWA-6, TWA-14, TWA-23), and the 7 targets belonging to the proposed Her-Lyr moving group (see Section~\ref{sec:sample}). \label{fig:mc3}}  
\end{figure*}

\subsubsection{$d\mathcal{N}/da$ distribution--independent}

We repeated all of the calculations presented in the previous section for fixed companion semimajor axes between 3--30\,AU. As before, when the imaging observations were included in the analysis, the upper limit estimates only change by $\sim$1--2\% over the entire 3--30\,AU depending on which assumption is made for the companion eccentricities. However, when the aperture masking observations are used on their own this variation increases to $\sim$5--10\% over the range 3--10\,AU and becomes as high as $\sim$20\% over the 10--30\,AU range (Figure~\ref{fig:mcecc}).

In Figure~\ref{fig:mc1} we plot the calculated upper limit estimates obtained for fixed companion masses and randomly-sampled companion masses, while holding the semimajor axes fixed at successive values between 3--30\,AU using a step size of 0.5\,AU and randomly drawing the companion eccentricities from a distribution of the form $d \mathcal{N} /de \propto 2e$ (Equation~\ref{eq:edist}). On their own, the aperture masking results place the tightest constraints over the $\sim$3--10\,AU semimajor axis range, with upper limit estimates of 20\%, 16\% and 13\% for fixed companion masses of 20\,\mj, 40\,\mj \ and 60\,\mj, respectively. With the imaging observations included in the analysis, these limits improve to 19\%, 13\% and 10\%, respectively. At larger separations between 10--30\,AU, our upper limit estimates are 12\%, 9\%, and 8\%, respectively, for the same companion masses when we include the imaging observations, while the companion frequencies are poorly constrained by the aperture masking observations alone.

Meanwhile, the right panel in Figure~\ref{fig:mc1} shows the results that were obtained when we sampled the companion masses from each of the three distributions given in Section~\ref{sec:mdist}. Over the 10--30\,AU semimajor axis range, the aperture masking observations on their own constrain the frequency of 20--80\,\mj \ companions to be less than 16\%, 15\% or 13\% at 95\% confidence, depending on whether the mass power law (Equation~\ref{eq:mpowerlaw}), mass ratio power law (Equation~\ref{eq:qpowerlaw}) or mass ratio log-normal paramaterization (Equation~\ref{eq:qlognormal}) is assumed for the companions. These constraints improve to 13\%, 12\% and 10\%, respectively, when the imaging observations are included in the analysis. At wider separations between 10--30\,AU, the equivalent values obtained when the aperture masking observations are combined with the previous imaging observations are 9\%, 9\% and 8\%, respectively.

Finally, Figures~\ref{fig:mc2} and \ref{fig:mc3} have been included for completeness. They are the same as Figure~\ref{fig:mc1} except that they show respectively the results obtained when upper values for the target ages and distances are used as described in Section~\ref{sec:mcresults1}, and the results obtained when the 9 targets of less certain membership identified in Section~\ref{sec:sample} and the Her-Lyr targets are not included in the calculations.

\subsection{Implications for Formation Theories} \label{sec:implications}

A well-known result from radial velocity surveys is the discovery of a ``brown dwarf desert'' at separations $\lesssim$3\,AU, where $\lesssim$0.5--1\% of solar-like stars are found to possess a 13--75\,\mj \ companion \citep{2000PASP..112..137M, 2006ApJ...640.1051G} compared with $\sim$10\% possessing a 0.3--10\,\mj \ companion \citep{2008PASP..120..531C} and $\sim$13\% possessing a $>$0.1\,\ms \ stellar companion \citep{1991A&A...248..485D}. Meanwhile, at wider separations, imaging surveys have started to place upper limits on the frequency of substellar companions:
\begin{itemize}
\item \cite{2006AJ....132.1146C} obtained a 95\% confidence upper limit of 12.1\% on the frequency of 13--73\,\mj \ companions between 25--100\,AU. 
\item \cite{2007ApJ...670.1367L} obtained a 95\% confidence interval of $1.9^{+8.3}_{-1.5}$\% for the frequency of 13--75\,\mj \ companions between 25--250\,AU.
\item \cite{2009ApJS..181...62M} obtained a 95\% confidence interval of $3.2^{+7.3}_{-2.7}$\% for the frequency of 13--75\,\mj \ companions between 28--1590\,AU. This is consistent with the results of \cite{2011ApJ...731....8K}, who obtained a lower bound of $3.9^{+2.6}_{-1.2}$\% for the frequency of substellar companions over the range 5--5000\,AU by combining the results of their aperture masking survey of Taurus-Auriga members with previous direct imaging results.
\item By jointly analyzing the results from three of the largest and deepest surveys for substellar companions to date \citep{2005ApJ...625.1004M, 2007ApJS..173..143B, 2007ApJ...670.1367L}, \cite{2010ApJ...717..878N} obtained 95\% confidence upper limits of $<$20\% and $<$5\% for the frequency of companions with masses between 10--15\,\mj \ in the ranges 13--600\,AU and 40--200\,AU, respectively.
\end{itemize}
The aperture masking survey reported in this paper has allowed us to place similar constraints on the frequency of 20--80\,\mj \ companions over the 3--30\,AU separation range (Sections~\ref{sec:mdist}--\ref{sec:adist}). 

These results are broadly in line with expectations from current models of substellar companion formation. Firstly, population synthesis models predict that core accretion only produces companions with masses up to $\sim$10\,\mj \ \citep{2004ApJ...604..388I}, or else, if objects are formed with masses above 20\,\mj, then these are extremely rare \citep{2009A&A...501.1139M}. Unsurprisingly, the observational studies outlined above provide no evidence to the contrary, despite the aperture masking surveys in particular probing the range of separations where core accretion is expected to be most efficient. 

Indeed, 20--80\,\mj \ companions are much more likely to form by either gravoturbulent fragmentation during the initial collapse of the molecular cloud \citep[eg.][]{2009MNRAS.392..590B, 2009ApJ...703..131O} or by the fragmentation of gravitational instabilities in the protostellar disk once the initial free-fall collapse of the molecular cloud has ended \citep[eg.][]{2009MNRAS.396.1066C, 2009MNRAS.392..413S}. For the gravoturbulent fragmentation scenario, the low frequencies of substellar companions deduced for separations $\lesssim$200\,AU from observational studies is in qualitative agreement with the hydrodynamical simulations of \cite{2009MNRAS.392..590B} who found that the separation of binary pairs consisting of a stellar primary and a very low-mass secondary increases strongly with decreasing mass ratio. Meanwhile, the disk fragmentation mechanism is not expected to occur within $\sim$40--70\,AU of the primary, where radiative cooling time scales are too long for the disk to be Toomre unstable \citep[eg.][]{2007ApJ...662..642R, 2009ApJ...695L..53B}. Alternatively, 20--80\,\mj \ objects might form by gravitational disk instabilities at separations beyond $\sim$40--70\,AU and then migrate inwards. \cite{2009MNRAS.392..413S} considered this for the case of a massive disk extending between 40--400\,AU around a 0.7\,\ms \ star. However, they found that when low mass ($<$80\,\mj) companions did form at closer separations, they were subsequently scattered outwards by dynamical interactions with more massive companions in the same disk, leading to a brown dwarf desert that extended out to $\sim$100--200\,AU. Again, the low occurence of 20--80\,\mj \ companions inferred from observational studies over this separation range is consistent with such a scenario, though the constraints are not yet tight enough to make a definitive statement.

\section{Conclusion} \label{sec:conclusion}

This paper has presented the results of an aperture masking survey of \ssize \ young nearby stars for substellar companions. Our detection limits extend down to $\sim$\,40\,\mj \ for 30 of our targets, and of these, we are sensitive down to $\sim$\,20\,\mj \ or less for a subset of 22.  Although we did not uncover any substellar companions, we detected four stellar companions. One of these, a $0.52 \pm 0.09$\,\ms \ companion to HIP\,14807, is a new discovery. We have also shown that the companion to HD\,16760 is on a low inclination orbit with a mass of $0.28 \pm 0.04$\,\ms, much higher than the minimum mass of $M_2 \, \sin i \sim $13--14\,\mj \ inferred from radial velocity measurements.

If we do not make any assumptions about the distribution of companion masses or semimajor axes, we calculate that the frequency of 20--80\,\mj \ companions is less than $\sim$\,19\% in the range 3--10\,AU and less than $\sim$\,12\% in the range 10--30\,AU at 95\% confidence. If, however, we assume that the semimajor axes of 20--80\,\mj \ companions are distributed according to $d\mathcal{N}/da \propto a^{-1}$ and that their masses are distributed according to a log-normal parameterization of the secondary-to-primary mass ratio, this limit becomes $\sim$9\% over the 3--30\,AU separation range. Similar values of $\sim$10\% and $\sim$11\% are obtained if we assume instead that the companion masses or secondary-to-primary mass ratios, respectively, are distributed according to power laws. These results are consistent with models that predict a low occurrence of substellar companions relative to stellar companions at these separations, possibly hinting at the extension of the brown dwarf desert beyond $\sim$\,3\,AU.

\acknowledgements

M.I. was the recipient of the Australian Research Council postdoctoral fellowship (project number DP0878674). A.K. was previously supported 
by a NASA/Origins grant to Lynne Hillenbrand and is currently supported by a NASA Hubble Fellowship grant. This work was also partially 
supported by the National Science Foundation under Grant Numbers 0506588 
and 0705085. This work made use of data products from 2MASS, which is a 
joint project of the University of Massachusetts and IPAC/Caltech, funded 
by NASA and the NSF. Our research has also made use of the USNOFS Image 
and Catalogue Archive operated by the United States Naval Observatory, 
Flagstaff Station (http://www.nofs.navy.mil/data/fchpix/).

We recognize and acknowledge the very significant cultural role and 
reverence that the summit of Mauna Kea has always had within the 
indigenous Hawaiian community. We are most fortunate to have the 
opportunity to conduct observations from this mountain.















\clearpage
\begin{deluxetable}{cccccc}
\vspace{4cm}
\tabletypesize{\scriptsize}
\tablewidth{0pt}
\tablecaption{Mask Properties \label{table:masks}}
\tablehead{
\colhead{Observatory} &
\colhead{Telescope} &
\colhead{Instrument} &
\colhead{Mask} &
\colhead{Hole Diameter} &
\colhead{Transmission} 
 \\
 \colhead{} & 
 \colhead{Diameter (m)} & 
 \colhead{} & 
\colhead{} &
\colhead{(m)}  &
\colhead{\%}
}
\startdata
Palomar & 5.1 &PHARO & 9H & 0.4 & 6 \\
            & && 18H & 0.2 & 3  \\
Keck & 10 &NIRC2  & 9H & 1.1 & 11\\
            & && 18H & 0.5 & 5 \\
VLT & 8.2 &CONICA & 7H & 1.2 & 15 \\
\enddata
\end{deluxetable}

\begin{deluxetable}{rrccc}
\vspace{2cm}
\tabletypesize{\scriptsize}
\tablewidth{0pt}
\tablecaption{Moving Groups Surveyed  \label{table:mg}}
\tablehead{
\multicolumn{1}{r}{Group} & 
\multicolumn{1}{r}{$N$\tablenotemark{a}} & 
\colhead{Age\,\tablenotemark{b}} &
\colhead{Distance\,\tablenotemark{c}} & 
\colhead{Reference\,\tablenotemark{d}} 
\\
\colhead{} & 
\colhead{} & 
\colhead{(Myr)} & 
\colhead{(pc)} & 
\colhead{}
}
\startdata
AB Dor & \ssizeabdor & $110 \pm 40$ & $34.1 \pm 12.8$ & Lu05, T08 \\
$\beta$ Pic & \ssizebpic & $12 \pm 5$ & $34.5 \pm 1.4$ & Z01a, F06, T08 \\ 
Her-Lyr & \ssizeherlyr & $200 \pm 80$ & $14.6 \pm 4.1$ & LS06 \\
Tuc-Hor & \ssizetuchor & $30 \pm 10$ & $45.1 \pm 0.6$ & Z01b, T08 \\
 TWA & \ssizetwa & $8 \pm 4$ & $55.0 \pm 2.7$ & dR06, T08 \\
\enddata
\tablenotetext{a}{The number of targets observed for each group in this survey.}
\tablenotetext{b}{Adopted age uncertainties are estimated from the spread in reported literature values and their associated uncertainties.}
\tablenotetext{c}{Quoted values and dispersions are the medians and standard deviations of those members observed in this survey (see Table~\ref{table:sample}).}
\tablenotetext{d}{\cite{2008hsf2.book..757T} (T08) provide a review of the memberships of four of the five groups targeted in this survey, while \cite{2006ApJ...643.1160L} (LS06) describe the proposed Her-Lyr association. Also cited here are \cite{2005ApJ...628L..69L} (Lu05), \cite{2006AJ....131.2609D} (dR06), \cite{2001ApJ...562L..87Z} (Z01a), \cite{2001ApJ...559..388Z} (Z01b), and \cite{2006AJ....131.1730F} (F06).}
\end{deluxetable}

\clearpage
\LongTables
\begin{landscape}
\begin{deluxetable*}{lrrccrcrrrccc} 
\tabletypesize{\scriptsize}
\tablecaption{Sample List \label{table:sample}}
\tablewidth{0pt}
\tablehead{
\colhead{Name} & 
\colhead{RA} & 
\colhead{DEC} &
\colhead{Spectral} & 
\colhead{Mass\,\tablenotemark{b}} & 
\colhead{Distance} & 
\colhead{Ref.\,\tablenotemark{c}} & 
\colhead{K} & 
\colhead{H} & 
\colhead{J} & 
\colhead{Ref.\,\tablenotemark{d}} &
\colhead{Membership} & 
\colhead{Previous}
\\
\colhead{} & 
\multicolumn{2}{c}{(J2000)} & 
\colhead{Type\,\tablenotemark{a}} &
\colhead{($M_\odot$)} &
\colhead{(pc)} & 
\colhead{} & 
\colhead{(mag)} & 
\colhead{(mag)} & 
\colhead{(mag)} & 
\colhead{} & 
\colhead{Ref.\,\tablenotemark{e}} & 
\colhead{\, Imaging\,\tablenotemark{f}}
}
\startdata
\multicolumn{11}{l}{AB Dor Targets} \\ 
\hline \\
PW And & 00 18 20.8 & +30 57 24 & K2 & 0.81  & 28 $\pm$ 7\,\tablenotemark{g} & ZS04 &6.39 & 6.51 & 7.02 & 2M & ZS04, T08 & MZ04, L05, L07, \\
& & & & & & & & & & & & MH09, H10 \\
HIP\,3589 & 00 45 50.9 & +54 58 40 & F8  & 1.12 & 52.5 $\pm$ 2.5 & HIP & 6.69 & 6.72 & 6.93 & 2M & ZS04, T08 & $\cdots$ \\
HIP\,5191 & 01 06 26.1 & $-14$ 17 46 & K1  & 0.86 & 47.3 $\pm$ 2.8 & HIP & 7.34 & 7.43 & 7.91 & 2M & ZS04, T08 & C10\\
HIP\,6276 & 01 20 32.2 & $-11$ 28 03 & G9  & 0.89 & 34.4 $\pm$ 1.2 & HIP & 6.55 & 6.65 & 7.03 & 2M & ZS04, T08 & MH09 \\
HIP\,12635 & 02 42 20.9 & +38 37 22 & K2  & 0.79 & 50.4 $\pm$ 6.7 & HIP & 7.76 & 7.90 & 8.38 & 2M & ZS04, T08 & $\cdots$ \\
HD\,16760 & 02 42 21.3 & +38 37 08 & G5  & 0.91 & 45.5 $\pm$ 4.9 & HIP & 7.11 & 7.15 & 7.47 & 2M & ZS04, T08 & $\cdots$  \\
HIP\,13027 & 02 47 27.4 & +19 22 19 & G1  & 1.02 & 33.6 $\pm$ 0.9 & HIP & 6.05 & 6.10 & 6.37 & 2M & ZS04, T08 & $\cdots$ \\ 
HD\,19668 & 03 09 42.3 & $-09$ 34 47 & G0  & 0.90 & 37.4 $\pm$ 1.6 & HIP & 6.70 & 6.79 & 7.16 & 2M & LS06, T08 & MH09 \\
HIP\,14807 & 03 11 12.3 & +22 25 24 & K6  & 0.76 & 52.5 $\pm$ 8.6 & HIP & 7.96 & 8.10 & 8.67 & 2M & ZS04, T08 & $\cdots$ \\
HIP\,14809 & 03 11 13.8 & +22 24 58 & G5  & 1.04 & 53.7 $\pm$ 3.3 & HIP & 6.97 & 7.07 & 7.27 & 2M & ZS04, T08 & $\cdots$ \\
HIP\,16563A & 03 33 13.5 & +46 15 27 & G5  & 0.88 & 34.4 $\pm$ 1.2 & HIP & 6.62 & 6.70 & 7.03 & 2M & ZS04, T08 & B07 \\
HIP\,16563B & 03 33 14.0 & +46 15 19 & M0  & 0.55 & 34.4 $\pm$ 1.2 & HIP & 8.07 & 8.21 & 8.83 & 2M & ZS04, T08 & $\cdots$ \\
HIP\,17695 & 03 47 23.2 & $-01$ 58 18 & M3  & 0.46 & 16.1 $\pm$ 0.7 & HIP& 6.93 & 7.17 & 7.80 & 2M & ZS04, T08 & L07 \\
HIP\,18859 & 04 02 36.7 & $-00$ 16 06 & F6  & 1.16 & 18.8 $\pm$ 0.1 & HIP & 4.18 & 4.34 & 4.71 & 2M & ZS04, T08 & L07, H10  \\
HIP\,19183 & 04 06 41.5 & +01 41 03 & F5  & 1.17  & 55.2 $\pm$ 2.8 & HIP & 6.58 & 6.70 & 6.89 & 2M & ZS04, T08 & $\cdots$  \\
BD+20 1790 & 07 23 44.0 & +20 25 06 & K5  & 0.76  & 32 $\pm$ 8\,\tablenotemark{g} & LS06 & 6.88 & 7.03 & 7.64 & 2M & LS06, T08  & L05, MH09, H10 \\
HD\,89744\,\tablenotemark{h, i} & 10 22 10.6 & +41 13 46 & F7  & 1.52 & 39.4 $\pm$ 0.5 & HIP & 4.45 & 4.53 & 4.86 & 2M  & LS06 & $\cdots$ \\
HIP\,51317 & 10 28 55.6 & +00 50 28 & M2  & 0.43 & 7.1 $\pm$ 0.1 & HIP & 5.31 & 5.61 & 6.18 & 2M & LS06, T08 & M05, L07 \\
HD\,92945\,\tablenotemark{h} & 10 43 28.3 & $-$29 03 51 & K1  & 0.85 & 21.4 $\pm$ 0.3 & HIP & 5.66 & 5.77 & 6.18 & 2M & LS06 & B07, L07 \\
GJ\,466\,\tablenotemark{h} & 12 25 58.6 & +08 03 44 & M0  & 0.73 & 37.4 $\pm$ 3.2 & HIP & 7.33 & 7.45 & 8.12 & 2M & LS06 & MZ04 \\
HD\,113449 & 13 03 49.8 & $-05$ 09 41 & K1 & 0.84 & 21.7 $\pm$ 0.4 & HIP & 5.72 & 5.89 & 6.27 & 2M & ZS04, T08 & L07, H10 \\
EK Dra\,\tablenotemark{h} & 14 39 00.2 & +64 17 30 & G1.5  & 1.06 & 34.1 $\pm$ 0.4 & HIP & 5.91 & 6.01 & 6.32 & 2M & LS06 & MZ04, B07, L07, \\
& & & & & & & & & & & & MH09 \\
HIP\,81084 & 16 33 41.7 & $-09$ 33 10 & M0  & 0.58  & 30.7 $\pm$ 2.3 & HIP & 7.55 & 7.78 & 8.38 & 2M & ZS04, T08 & L07 \\
HIP\,82688 & 16 54 08.2 & $-04$ 20 24 & G0  & 1.12  & 46.7 $\pm$ 2.0 & HIP & 6.36 & 6.48 & 6.70 & 2M & ZS04, T08 &  MH09 \\
HD\,160934 & 17 38 39.7 & +61 14 16 & K7  & 0.70  & 33.1 $\pm$ 2.2 & HIP & 7.22 & 7.37 & 7.98 & 2M & ZS04, T08 &  L05, H07, L07, \\
& & & & & & & & & & & & MZ04 \\
HIP\,106231 & 21 31 01.6 & +23 20 09 & K5  & 0.75  & 24.8 $\pm$ 0.7 & HIP & 6.38 & 6.52 & 7.08 & 2M & ZS04, T08 & L05, L07, MZ04  \\
HIP\,110526 & 22 23 29.1 & +32 27 34 & M3  & 0.48  & 15.5 $\pm$ 1.6 & HIP & 6.05 & 6.28 & 6.90 & 2M & ZS04, T08 & $\cdots$ \\
HIP\,113579 & 23 00 19.2 & $-26$ 09 12 & G5  & 0.99 & 30.8 $\pm$ 0.7 & HIP & 5.94 & 6.04 & 6.29 & 2M & ZS04, T08 & MH09, C10  \\
HIP\,114066 & 23 06 04.6 & +63 55 35 & M1  & 0.60  & 24.5 $\pm$ 1.0 & HIP & 6.98 & 7.17 & 7.82 & 2M & ZS04, T08 & L07 \\
HIP\,115162 & 23 19 39.5 & +42 15 10 & G4  & 0.94  & 50.2 $\pm$ 2.9 & HIP & 7.22 & 7.28 & 7.61 & 2M & ZS04, T08 & $\cdots$  \\
HIP\,118008 & 23 56 10.5 & $-39$ 03 07 & K2  & 0.81 & 22.0 $\pm$ 0.4 & HIP & 5.91 & 6.01 & 6.51 & 2M & ZS04, T08 & B07, C10 \\

\\
\multicolumn{11}{l}{$\beta$ Pic Targets} \\ 
\hline \\
HR9 & 00 06 50.1 & $-23$ 06 27 & F3  & 1.40 & 39.4 $\pm$ 0.6 & HIP & 5.24 & 5.33 & 5.45 & 2M & ZS04, T08 & K03 \\ 
HIP\,10680 & 02 17 25.2 & +28 44 43 & F5  & 1.15 & 34.5 $\pm$ 0.6 & HIP & 5.79 & 5.84 & 6.05 & 2M & ZS04, T08 & $\cdots$ \\ 
HIP\,11437B & 02 27 28.1 & +30 58 41 & M2  & 0.35  & 40.0 $\pm$ 3.6 & HIP & 7.92 & 8.14 & 8.82 & 2M & ZS04, T08 & $\cdots$  \\ 
HIP\,11437A & 02 27 29.2 & +30 58 25 & K6  & 0.63  & 40.0 $\pm$ 3.6 & HIP & 7.08 & 7.24 & 7.87 & 2M & ZS04, T08 & $\cdots$ \\ 
HIP\,12545 & 02 41 25.8 & +05 59 19 & K6  & 0.67 & 42.0 $\pm$ 2.7 & HIP & 7.07 & 7.23 & 7.9 & 2M & ZS04, T08 & B07 \\ 
51 Eri & 04 37 36.1 & $-02$ 28 25 & F0  & 1.41 & 29.4 $\pm$ 0.3 & HIP & 4.54 & 4.77 & 4.74 & 2M & ZS04, T08 & H10 \\ 
HIP\,25486 & 05 27 04.8 & $-11$ 54 04 & F7  & 1.25 & 27.0 $\pm$ 0.4 & HIP & 4.93 & 5.09 & 5.27 & 2M & ZS04, T08 & L05, K07, MH09  \\ 
GJ\,803 & 20 45 09.5 & $-31$ 20 27 & M1  & 0.44 & 9.9 $\pm$ 0.1 & HIP & 4.53 & 4.83 & 5.44 & 2M & ZS04, T08 & K03, MZ04, M05, \\
& & & & & & & & & & & & B07, L07 \\ 
BD$-$17 6128 & 20 56 02.7 & $-17$ 10 54 & K6  & 0.73  & 45.7 $\pm$ 1.6 & HIP & 7.12 & 7.25 & 7.92 & K04 & ZS04, T08 & M05 \\ 
HIP\,112312A & 22 44 57.9 & $-33$ 15 02 & M4  & 0.31  & 23.3 $\pm$ 2.0 & HIP & 6.93 & 7.15 & 7.79 & 2M & ZS04, T08 & B07  \\ 
HIP\,112312B & 22 45 00.0 & $-33$ 15 26 & M4.5  & 0.17 & 23.3 $\pm$ 2.0 & HIP & 7.79 & 8.06 & 8.68 & 2M & ZS04, T08 & $\cdots$ \\ 

\\         
\multicolumn{11}{l}{Her-Lyr Targets} \\
\hline  \\
HD\,166 & 00 06 36.8 & 29 01 17.4 & K0  & 0.93 & 13.7 $\pm$ 0.1 & HIP & 4.31 & 4.63 & 4.73 & 2M & LS06 & L07, H10 \\
HD\,10008 & 01 37 35.5 & -06 45 37.5 & G5  & 0.89 & 24.0 $\pm$ 0.4 & HIP & 5.75 & 5.90 & 6.23 & 2M & LS06 & L07 \\
HD\,233153 & 05 41 30.7 & +53 29 23 & M0.5  & 0.58 & 12.4 $\pm$ 0.3 & HIP & 5.76 & 5.96 & 6.59 & 2M & LS06 & C05 \\
HIP\,37288 & 07 39 23.0 & +02 11 01 & K7  & 0.61  & 14.6 $\pm$ 0.3 & HIP & 5.87 & 6.09 & 6.77 & 2M & LS06 & M05, L07 \\
HD\,70573 & 08 22 50.0 & 01 51 33.6 & G6  & 0.89  & 46 $\pm$ 11\,\tablenotemark{g} & LS06 & 7.19 & 7.28 & 7.56 & 2M & LS06 & L05, MH09 \\
HIP\,53020 & 10 50 52.1 & +06 48 29 & M4  & 0.25  & 6.8 $\pm$ 0.2 & HIP & 6.37 & 6.71 & 7.32 & 2M & LS06 & L07 \\
HN Peg & 21 44 31.3 & +14 46 19 & G0  & 1.06 & 17.9 $\pm$ 0.1 & HIP & 4.56 & 4.6 & 4.79 & 2M & LS06 & MZ04, L07  \\

\\
\multicolumn{11}{l}{Tuc-Hor Targets} \\
\hline \\
HIP\,9141 & 01 57 48.9 & $-21$ 54 05 & G4  & 0.97 & 40.9 $\pm$ 1.1 & HIP & 6.47 & 6.56 & 6.86 & 2M & ZS04, T08 & B07, MH09  \\
HIP\,30030\,\tablenotemark{h} & 06 19 08.1 & $-03$ 26 20 & G0  & 1.03 & 49.2 $\pm$ 2.0 & HIP & 6.55 & 6.59 & 6.85 & 2M & ZS04 & B07, MH09 \\

\\
\multicolumn{11}{l}{TWA Targets} \\ 
\hline \\
TWA-21\,\tablenotemark{h} & 10 13 14.8 & $-52$ 30 54 & K3/4  & 0.63 & 48 $\pm$ 4 & MM05 & 7.19 & 7.35 & 7.87 & 2M & ZS04 & $\cdots$ \\ 
TWA-6\,\tablenotemark{h} & 10 18 28.8 & $-31$ 50 02 & K7  & 0.43 & 55 $\pm$ 5 & MM05 & 8.04 & 8.18 & 8.87 & 2M & ZS04 & W99, MZ04, L05, \\
& & & & & & & & & & & & M05 \\  
TWA-7 & 10 42 30.3 & $-33$ 40 17 & M2  & 0.35 & 29 $\pm$ 2 & MM05 & 6.9 & 7.13 & 7.79 &  2M & ZS04, T08 & W99, MZ04, L05  \\ 
TW Hya & 11 01 51.9 & $-34$ 42 17 & K6  & 0.64 & 53.7 $\pm$ 6.2 & HIP & 7.30 & 7.56 & 8.22 & 2M & ZS04, T08 & W99, MZ04, L05  \\ 
TWA-3 & 11 10 28.0 & $-37$ 31 53 & M4  & 0.37 & 36 $\pm$ 4 & MM05 & 7.28 & 7.60 & $\cdots$ &  W00 & ZS04, T08 & W99, C10 \\ 
TWA-14\,\tablenotemark{h} & 11 13 26.5 & $-45$ 23 43 & M0  & 0.57 & 86 $\pm$ 8 & MM05 & 8.50 & 8.73 & 9.42 &  2M & ZS04 & B07, MZ04, C10   \\ 
TWA-13B & 11 21 17.2 & $-34$ 46 45 & M1  & 0.63  & 57 $\pm$ 10 & MM05 & 7.49 & 7.73 & 8.43 &  2M & ZS04, T08 & $\cdots$ \\ 
TWA-13A & 11 21 17.5 & $-34$ 46 50 & M1  & 0.61  & 57 $\pm$ 10 & MM05 & 7.46 & 7.68 & 8.43 &  2M & ZS04, T08 & $\cdots$ \\ 
TWA-8B & 11 32 41.4 & $-26$ 52 08 & M5  & 0.14  & 42 $\pm$ 5 & MM05 & 9.01 & 9.36 & $\cdots$ &  W00 & ZS04, T08 & W99, M05, L05  \\ 
TWA-8A & 11 32 41.5 & $-26$ 51 55 & M3  & 0.40  & 41 $\pm$ 4 & MM05 & 7.44 & 7.72 & $\cdots$ &  W00 & ZS04, T08 & W99, M05 \\ 
TWA-9 & 11 48 24.2 & $-37$ 28 49 & K5  & 0.38 & 46.8 $\pm$ 5.4 & HIP & 7.85 & 8.03 & 8.68 & 2M & ZS04, T08 & W99, M05 \\ 
TWA-23\,\tablenotemark{h} & 12 07 27.4 & $-32$ 47 00 & M1  & 0.58 & 61 $\pm$ 5 & MM05 & 7.75 & 8.03 & 8.62 &  2M & ZS04 & C10 \\ 
TWA-25 & 12 15 30.8 & $-39$ 48 42 & M1  & 0.68  & 55 $\pm$ 4 & MM05 & 7.31 & 7.50 & 8.17 &  2M & ZS04, T08 & B07, C10 \\ 
TWA-10 & 12 35 04.3 & $-41$ 36 39 & M2  & 0.39  & 57 $\pm$ 9 & MM05 & 8.19 & 8.48 & 9.12 &  2M & ZS04, T08 & W99, MZ04, L05  \\ 
TWA-11B & 12 36 00.6 & $-39$ 52 16 & M2  & 0.52  & 72.8 $\pm$ 1.7 & HIP & 8.35 & 8.53 & 9.15 & 2M & ZS04, T08 & W99  \\ 
TWA-11A & 12 36 01.0 & $-39$ 52 10 & A0  & 2.31  & 72.8 $\pm$ 1.7 & HIP & 5.77 & 5.79 & 5.78 & 2M & ZS04, T08 & W99, C10 \enddata
\tablenotetext{a}{Spectral types are taken from the lists contained in \cite{2004ARA&A..42..685Z}, \cite{2006ApJ...643.1160L}, and \cite{2008hsf2.book..757T}.}
\tablenotetext{b}{Masses are estimated by interpolation of the NextGen isochrones of \cite{1998A&A...337..403B}, assuming the ages listed in Table~\ref{table:mg}. The only exceptions are HD\,89744 and TWA-11A which fall outside the range of the NextGen isochrones. For these two targets, we estimated their masses using the Y${}^2$ isochrones of \cite{2001ApJS..136..417Y}. Allowing for the uncertainties in the distances and ages, the calculated masses are typically uncertain at the $\la$5--10\% level for the AB Dor targets, $\la$30\% for the $\beta$ Pic targets, $\la$5\% for the Her-Lyr and Tuc-Hor targets, and $\la$40\% for the TWA targets. These values, however, do not account for uncertainties in the isochrones themselves, nor for uncorrected blending.}
\tablenotetext{c}{Distances have been obtained from HIPPARCOS (HIP) parallaxes where available \citep{2007ASSL..350.....V}. In the case of BD$-$17 6128, the HIPPARCOS distance to HD\,199143 is quoted, since  \cite{2000A&A...363L..25V} have shown that they form a physical pair. Otherwise distances have been derived photometrically, as listed by \cite{2004ARA&A..42..685Z} (ZS04), \cite{2006ApJ...643.1160L} (LS06), and \cite{2005ApJ...634.1385M} (MM05). Note that here the MM05 distances have been increased upwards by 7\% following the revision of \cite{2007ApJ...668L.175M}. }
\tablenotetext{d}{2MASS (2M) photometry is used where available \citep{2003tmc..book.....C}. A blend correction is applied in cases where the target is known to be binary but is unresolved in the 2MASS catalog. For other targets, quoted magnitudes were obtained from \cite{2004A&A...414..175K} (K04) and \cite{2000ApJ...530..867W} (W00).}
\tablenotetext{e}{\cite{2004ARA&A..42..685Z} (ZS04), \cite{2006ApJ...643.1160L} (LS06), and \cite{2008hsf2.book..757T}. A number of targets were originally on the lists of ZS04 or proposed by LS06, but not confirmed by T08.}
\tablenotetext{f}{Previous high-resolution imaging surveys that have observed the target. Most used coronagraphic or differential imaging techniques to search for low-mass companions at wider separations that are complementary to ours. \cite{2007ApJS..173..143B} (B07), \cite{2005AJ....130.1212C} (C05), \cite{2010A&A...509A..52C} (C10), \cite{2007AA...463..707H} (H07),  \cite{2010ApJ...714.1551H, 2010ApJ...714.1570H} (H10), \cite{2003ASPC..294...91K} (K03), \cite{2007A&A...472..321K} (K07), \cite{2005AJ....130.1845L} (L05), \cite{2007ApJ...670.1367L} (L07), \cite{2005ApJ...625.1004M} (M05),  \cite{2009ApJS..181...62M} (MH09), \cite{2004AJ....127.2871M} (MZ04), \cite{1999ApJ...512L..63W} (W99).}
\tablenotetext{g}{Uncertainty of 25\% assumed for photometrically-determined distances.}
\tablenotetext{h}{Proposed members that were not confirmed by the dynamical convergence method of T08. The analysis described in Section~\ref{sec:mc} was repeated separately both with and without these 9 targets included.}
\tablenotetext{i}{HD\,89744 is orbited by a substellar companion ($M_2 \sin i \sim $7--8\,\mj) with a 256 day period \citep{2000ApJ...533L.147K, 2006ApJ...646..505B}. Isochrone fits suggest that the system is $\sim$2\,Gyr old \citep{1998A&A...329..943N, 2001AJ....121..432G}, and hence does not belong to the AB Dor moving group, as suggested by LS06. This older age is supported by evidence from gyrochronology \citep{2007ApJ...669.1167B}. Furthermore, T08 did not find strong evidence for HD\,89744 being a member of AB Dor on dynamical grounds. However, we have chosen to retain HD\,89744 in our sample for two reasons: (1)~with a semimajor axis of 0.88\,AU and a distance of 39.4\,pc, the companion was potentially within our range of detectability; however, (2)~in the end we were only sensitive to companions with masses $>$200\,\mj \ around HD\,89744 (Table~\ref{table:detlims}), and as such, its inclusion does not affect the calculations presented in Section \ref{sec:analysis}, as these only included targets with sensitivity limits extending below 80\,\mj.}
\end{deluxetable*}
\clearpage
\end{landscape}

\begin{deluxetable}{ccccc}
\vspace{3cm}
\tabletypesize{\scriptsize}
\tablewidth{0pt}
\tablecaption{Filters Used \label{table:filters}}
\tablehead
{
\colhead{Instrument} &
\colhead{Filter} &
\colhead{$\lambda_0$ ($\mu$m)} &
\colhead{$\Delta\lambda$ ($\mu$m) }
}
\startdata
PHARO  &CH4s & 1.57 & 0.10 \\
 & H  & 1.64 & 0.30 \\
 & Ks  & 2.15 & 0.31 \\
\hline NIRC2   &Jcont  & 1.21 & 0.02 \\
 & Hcont & 1.58  & 0.02 \\
 & CH4s  & 1.59 & 0.13 \\
 & Kp  & 2.12 & 0.35 \\
 & Kcont  & 2.27 & 0.03 \\
 & CO  & 2.29 & 0.03 \\
\hline CONICA  & L${}^\prime$  & 3.80  & 0.62  \\
\enddata
\end{deluxetable}

\vspace{3cm}
\LongTables
\begin{deluxetable}{lccrcrc}
\tabletypesize{\scriptsize}
\tablewidth{0pt}
\tablecaption{Observations \label{table:observations}}
\tablehead{\colhead{Target} & 
\colhead{Instrument} &
\colhead{Filter} & 
\colhead{Mask} & 
\colhead{$T_{\rm int}$ (min)} &
\colhead{Date\,\tablenotemark{a}} &
\colhead{MJD} 
}
\startdata
\multicolumn{5}{l}{AB Dor Targets} \\
\hline \\

PW And  &  PHARO  & CH4s &  18H  & 3.9 & 2007 Jun 01 &  54252.5  \\ 

HIP\,3589  &  PHARO  & CH4s & 9H  & 6.5 & 2007 Nov 27 &  54431.1  \\

HIP\,5191  &  NIRC2  & Hcont & 18H  & 4.0 & 2007 Jun 06  &  54257.6  \\ 

HIP\,6276  &  NIRC2  & Kcont & 18H  & 5.3 & 2007 Nov 23 &  54427.2  \\ 

HIP\,12635  &  NIRC2  & Kcont & 18H  & 800 & 2008 Dec 23 &  54823.2  \\ 

HD\,16760\tablenotemark{b} & NIRC2 & Kcont & 18H  & 2.7 & 2008 Dec 23 & 54823.2 \\ 

& NIRC2 & Kcont & 18H  & 13 &  2009 Aug 06 & 55049.6  \\ 

& NIRC2 & Jcont & 18H  & 2.7 & 2009 Nov 20 & 55155.3 \\ 

& NIRC2 & Hcont & 18H  & 2.7 & 2009 Nov 20 & 55155.3 \\ 

& NIRC2 & Kcont & 18H  & 2.7 & 2009 Nov 20 & 55155.3  \\ 

& NIRC2 & CO & 18H  & 2.7 & 2009 Nov 21 & 55156.2 \\ 

HIP\,13027 & PHARO & Ks & 9H & 19 & 2007 Nov 27 & 54431.2 \\ 

HD\,19668 & PHARO  & Ks & 9H  & 5.8 & 2007 Nov 29 &  54433.3  \\ 

HIP\,14807\tablenotemark{b} & PHARO & Ks & 9H & 5.8 & 2007 Nov 29 & 54433.1 \\ 

& NIRC2 & CO & 18H & 5.0 & 2009 Nov 21 & 55156.2 \\

HIP\,14809  &  NIRC2  & CO & 18H & 2.7  & 2009 Nov 21 &  55156.2  \\ 

HIP\,16563A  &  PHARO  & Ks & 9H  & 5.8 & 2007 Nov 27 &  54431.3  \\ 

HIP\,16563B  &  NIRC2  & CO & 18H & 5.3  & 2009 Nov 21 &  55156.2  \\

HIP\,17695  &  PHARO  & Ks & 9H  & 12 & 2007 Nov 29 &  54433.3  \\ 

HIP\,18859  &  PHARO  & Ks & 9H  & 8.7 & 2007 Nov 27 &  54431.2  \\ 

HIP\,19183  &  PHARO  & Ks & 9H  & 8.7 & 2007 Nov 27 &  54431.3  \\ 

BD+20 1790  &  PHARO  & Ks & 9H  & 5.8 & 2007 Nov 27 &  54431.5  \\ 

HD\,89744     &  PHARO  & CH4s & 18H & 8.6  & 2007 Apr 05 &  54195.3  \\ 

HIP\,51317  &  PHARO  & Ks & 9H  & 13 &  2007 Apr 06 &  54196.3  \\ 

HD\,92945  &  NIRC2  & Kcont & 18H  & 1.3 & 2007 Jun 06 &  54257.2  \\ 

GJ\,466  &  PHARO  & Ks & 9H  & 2.2 & 2008 Jun 19 &  54636.2  \\ 

HD\,113449\tablenotemark{b} & PHARO & CH4s & 18H  & 13 & 2007 Apr 06 & 54196.3 \\ 

& PHARO & Ks & 9H  & 7.5 & 2007 Apr 07 & 54197.4 \\

& PHARO & CH4s & 18H  & 3.9 & 2007 Jun 01 & 54252.1 \\

& NIRC2 & Hcont & 18H  & 1.3 & 2008 Jun 17 & 54634.2 \\

& NIRC2 & CH4s & 9H  & 1.7 & 2008 Dec 21 & 54821.7 \\ 

& NIRC2 & Hcont & 18H & 2.7 & 2010 Apr 25 & 55311.4 \\ 

EK Dra  &  PHARO  & Ks & 9H  & 4.3 & 2008 Jun 20 &  54637.2  \\ 

HIP\,81084  &  PHARO  & CH4s & 18H  & 4.3 & 2007 May 30 &  54250.2  \\

HIP\,82688  &  PHARO  & CH4s & 18H  & 4.3 & 2007 Jun 02 &  54253.4  \\

HD\,160934\tablenotemark{b} & PHARO & H & 9H  & 9.7 & 2008 Jun 23 & 54640.3 \\ 

 & PHARO & Ks & 9H  & 8.6 & 2008 Jun 23 & 54640.3 \\ 
 
 & NIRC2 & Kcont & Clear\,\tablenotemark{c} & 0.2 & 2010 Apr 26 & 55312.6 \\

& NIRC2 & Jcont & 18H & 2.7 & 2011 Apr 23 & 55674.6 \\ 
 & NIRC2 & Hcont & 18H & 2.7 & 2011 Apr 23 & 55674.6 \\ 

HIP\,106231  &  PHARO  & CH4s & 18H & 4.8 & 2007 May 31 & 54251.4  \\ 

HIP\,110526  &  PHARO  & CH4s & 9H  & 1.9 & 2007 May 31 &  54251.4  \\ 

HIP\,113579  &  NIRC2  & Hcont & 18H  & 2.7 & 2007 Jun 05 &  54256.6  \\ 

HIP\,114066  &  PHARO  & CH4s & 18H  & 3.9 & 2007 Jun 01 &  54252.4 \\ 

HIP\,115162  &  NIRC2  & Kcont & 18H  & 5.3 & 2009 Nov 21 &  55156.2  \\ 

HIP\,118008  &  NIRC2  & Kcont & 18H  & 5.3 & 2007 Nov 23 &  54427.2  \\ 

\\
\multicolumn{5}{l}{$\beta$ Pic Targets} \\
\hline \\

HR9  &  NIRC2  & Hcont & 18H  & 2.7 & 2007 Jun 06 &  54257.6  \\

HIP\,10680  &  NIRC2 & Kcont &  18H  & 5.3 & 2007 Nov 24 &  54428.4  \\

HIP\,11437A  &  NIRC2  & Kcont & 18H & 5.3 & 2007 Nov 24 &  54428.4  \\

HIP\,11437B  &  NIRC2  & Kcont & 18H & 5.3 & 2007 Nov 24 &  54428.4  \\

HIP\,12545  &  NIRC2  & Kcont & 18H  & 5.3 & 2007 Nov 24 &  54428.4  \\

51 Eri  &  NIRC2  & Kcont & 18H & 5.3 & 2008 Dec 21 &  54821.4  \\

HIP\,25486  &  PHARO  & Ks & 9H & 12 & 2007 Nov 27 &  54431.4  \\

GJ\,803  &  NIRC2  & Hcont & 18H & 4.0 & 2007 Jun 05 &  54256.5  \\ 

BD$-$17 6128  &  PHARO & CH4s &  9H  & 4.3 & 2007 May 30 &  54250.5  \\ 

HIP\,112312A  &  NIRC2  & Kp & 9H  & 10 & 2008 Jun 17 &  54634.6  \\ 

HIP\,112312B  &  NIRC2  & Hcont & 18H  & 2.7 & 2007 Jun 05 &  54256.6  \\

\\
\multicolumn{5}{l}{Her-Lyr Targets} \\
\hline  \\

HD\,166 & PHARO & CH4s & 18H  & 5.8 & 2007 May 31 & 54251.5 \\

HD\,10008 & NIRC2 & Kcont & 18H  & 5.3 & 2007 Nov 23 & 54427.3 \\

HD\,233153  &  PHARO  & Ks & 9H  & 8.7 & 2007 Nov 27 & 54431.5  \\

HIP\,37288  &  NIRC2  & Kcont & 18H  & 5.3 & 2007 Nov 24 & 54428.6  \\ 

HD\,70573 & PHARO & Ks & 9H  & 13 & 2007 Nov 27 & 54431.5 \\ 

HIP\,53020  &  NIRC2  & Kcont & 18H  & 2.3 & 2007 Jun 06 &  54257.3  \\ 

HN Peg  &  PHARO  & CH4s & 18H  & 3.9 & 2007 Jun 01 &  54252.5  \\ 

\\
\multicolumn{5}{l}{Tuc-Hor Targets} \\
\hline \\                  

HIP\,9141  &  PHARO  & Ks & 9H  & 9.7 & 2007 Nov 29 &  54433.2  \\

HIP\,30030  &  NIRC2  & Kcont & 18H  & 5.3 & 2007 Nov 24 &  54428.6     \\

\\
\multicolumn{5}{l}{TWA Targets} \\
\hline \\

TWA-21  &  CONICA  & L${}^\prime$ & 7H  & 20 & 2009 Mar 06 &  54896.1  \\ 

TWA-6  &  CONICA  & L${}^\prime$ &  7H  & 20 & 2009 Mar 07 &  54897.1  \\ 

TWA-7  &  CONICA  & L${}^\prime$ &  7H  & 40 & 2009 Mar 06 &  54896.1  \\ 

TW Hya  &  CONICA  & L${}^\prime$ &  7H  & 60 & 2009 Mar 07 &  54897.2  \\ 

TWA-3  &  NIRC2  & Kp & 9H  & 11 & 2008 Dec 22 & 54822.7  \\ 

TWA-14  &  CONICA  & L${}^\prime$ &  7H  & 20 & 2009 Mar 07 &  54897.1  \\ 

TWA-13B  &  NIRC2  & Kp &  9H  & 5.3 & 2008 Dec 21 &  54821.6  \\ 

TWA-13A  &  CONICA  & L${}^\prime$ &  7H  & 40 & 2009 Mar 06 &  54896.2  \\ 

TWA-8B  &  NIRC2  & Kp &  9H  & 5.3 & 2008 Dec 23 &  54823.7  \\ 

TWA-8A  &  CONICA  & L${}^\prime$ &  7H  & 20  & 2009 Mar 05 &  54895.3  \\ 

TWA-9  &  CONICA  & L${}^\prime$ &  7H  & 20 & 2009 Mar 05 &  54895.3  \\ 

TWA-23  &  CONICA  & L${}^\prime$ &  7H  & 25 & 2009 Mar 05 &  54895.3  \\

TWA-25  &  CONICA  & L${}^\prime$ &  7H  & 20 & 2009 Mar 06 &  54896.2  \\ 
 
TWA-10  &  CONICA  & L${}^\prime$ &  7H  & 20 & 2009 Mar 06 &  54896.3  \\ 

TWA-11B  &  CONICA  & L${}^\prime$ &  7H  & 20 & 2009 Mar 07 &  54897.3  \\ 

TWA-11A  &  CONICA  & L${}^\prime$ &  7H  & 50 & 2009 Mar 07 &  54897.4  \\
\enddata
\tablenotetext{a}{For those targets that were observed on multiple epochs, we only report the observation that gave the deepest limits. The only exceptions are for those targets with detected companions, for which all epochs are reported. }
\tablenotetext{b}{Companion detected in the current survey.}
\tablenotetext{c}{Observation made in imaging mode without aperture mask.}
\end{deluxetable}

\clearpage
\begin{landscape}
\begin{deluxetable*}{lclccccccc}
\tabletypesize{\scriptsize}
\tablewidth{0pt}
\tablecaption{Detected Companions \label{table:binariesdetected}}
\tablehead{
\colhead{Primary} & 
\colhead{Instrument} & 
\colhead{Date} &
\colhead{MJD} &
\colhead{Filter} & 
\colhead{$\lambda_0$ ($\mu m$)} & 
\colhead{$\rho$ (mas)} &
\colhead{$\theta$ (deg)} &
\colhead{$C$} &
\colhead{$M_2$ ($M_\odot$)\,\tablenotemark{a}} 
}
\startdata
\\
HIP14807        & PHARO & 2007 Nov 29 &54433.1\,\tablenotemark{b}  & Ks & 2.15 & $63.22$            & $246.86$          & $10.15$           & $\cdots$          \\ 
                & NIRC2 & 2009 Nov 21 &55156.2                     & CO & 2.29 & $28.74 \pm 0.19$   & $89.74 \pm 0.29$  & $3.00 \pm 0.06$   & $0.52 \pm 0.09$ \\ 
\\

\hline \\ 
HD16760         & NIRC2 & 2008 Dec 23 &54823.2                      & Kcont & 2.27  & $26.11 \pm 2.59$  & $46.20 \pm 1.26$  & $13.48 \pm 3.15$      & $0.32 \pm 0.11$ \\ 
                & NIRC2 & 2009 Aug 6 &55049.6                      & Kcont & 2.27  & $26.78 \pm 0.90$  & $204.54 \pm 0.45$ & $13.11 \pm 1.00$       & $0.32 \pm 0.09$ \\  
                & NIRC2 & 2009 Nov 20 &55155.3                      & Jcont & 1.21  & $28.13 \pm 1.93$  & $286.50 \pm 3.62$ & $31.04 \pm 5.92$      & $0.24 \pm 0.08$ \\ 
                & NIRC2 & 2009 Nov 20 &55155.3                      & Hcont & 1.58  & $26.06 \pm 1.75$  & $286.87 \pm 1.94$ & $20.53 \pm 1.88$      & $0.27 \pm 0.08$ \\ 
                & NIRC2 & 2009 Nov 20 &55155.3\,\tablenotemark{b}   & Kcont & 2.27  & $39.07$           & $286.67$          & $26.58$               & $\cdots$ \\  
                & NIRC2 & 2009 Nov 21 &55156.2                      & CO & 2.29  & $25.37 \pm 4.51$  & $290.25 \pm 1.95$ & $15.18 \pm 6.44$         & $0.29 \pm 0.15$ \\  

\\

\hline \\ 
HD113449        & PHARO & 2007 Apr 6 &54196.3                     & CH4s & 1.57  & $35.62 \pm 0.51$  & $225.19 \pm 0.44$ & $4.27 \pm 0.25$      & $0.51 \pm 0.02$ \\  
                & PHARO & 2007 Apr 7 &54197.4\,\tablenotemark{c}  & Ks  & 2.15  & $40.46$ & $223.30$ & $6.68$      & $\cdots$ \\  
                & PHARO & 2007 Jun 1 &54252.1\,\tablenotemark{b}  & CH4s& 1.57  & $28.46$           & $179.92$          & $13.73$               & $\cdots$ \\ 
                & NIRC2 & 2008 Jun 17 &54634.2                     & Hcont  & 1.58  & $36.68 \pm 0.13$  & $222.91 \pm 0.21$ & $4.65 \pm 0.05$        & $0.50 \pm 0.02$ \\  
                & NIRC2 & 2008 Dec 21 &54821.7                     & CH4s & 1.59  & $27.87 \pm 0.13$  & $250.14 \pm 0.18$ & $4.62 \pm 0.03$      & $0.51 \pm 0.03$ \\  
                & NIRC2 & 2010 Apr 25 &55311.4                     & Hcont  & 1.58  & $35.81 \pm 0.17$  & $202.38 \pm 0.23$ & $4.58 \pm 0.06$        & $0.51 \pm 0.02$ \\  
 \\

\hline \\ 
HD160934       & PHARO & 2008 Jun 23 &54640.3                      & H   & 1.64  & $169.24 \pm 0.13$  & $273.35 \pm 0.05$ & $2.22\pm0.01$             & $0.54 \pm 0.03$ \\ 
               & PHARO & 2008 Jun 23 &54640.3                      & Ks  & 2.15  & $169.79 \pm 0.25$  & $273.29 \pm 0.09$ & $2.11 \pm 0.02$           & $0.54 \pm 0.04$ \\ 
               & NIRC2 & 2010 Apr 26 &55312.6\,\tablenotemark{d}   & Kcont  & 2.27  & $68.8 \pm 0.7$     & $290.0 \pm 0.6$   & $2.1 \pm 0.2$          & $0.54 \pm 0.04$ \\
               & NIRC2 & 2011 Apr 23 &55674.6                      & Jcont  & 1.21  & $19.96 \pm 0.05$   & $18.44 \pm 0.12$  & $2.21 \pm 0.01$        & $0.54 \pm 0.03$ \\
               & NIRC2 & 2011 Apr 23 &55674.6                      & Hcont  & 1.58  & $20.00 \pm 0.03$   & $18.42 \pm 0.09$  & $2.18 \pm 0.01$        & $0.54 \pm 0.03$ \\
\enddata
\tablenotetext{a}{The quoted masses were calculated by interpolation of the NextGen models of \cite{1998A&A...337..403B}, using the primary magnitude in the approriate band as listed in Table~\ref{table:sample} and the best-fit contrast~$C$. Given that all of our companions were detected around AB~Dor targets, we assumed a system age of $110\pm 40$\,Myr (Table~\ref{table:mg}). Quoted uncertainties were calculated the same way as described in the footnotes of Table~\ref{table:sample}. The uncertainty associated with each mass value is the quadrature sum of three components:~(1) $\Delta M_C = |M(C-\sigma_C)-M(C+\sigma_C)|/2$, where $C$ is the fitted contrast with associated uncertainty $\sigma_C$; (2) $\Delta M_Y = |M(Y-\sigma_Y)-M(Y+\sigma_Y)|/2$, where $Y$ is the age with associated uncertainty $\sigma_Y$ as listed in Table~\ref{table:mg}; and (3) $\Delta M_d = |M(d-\sigma_d)-M(d+\sigma_d)|/2$, where $d$ is the distance with associated uncertainty $\sigma_d$ as listed in Column 6 of Table~\ref{table:sample}.}
\tablenotetext{b}{Due to a contrast/separation degeneracy at close separations, these fits were unreliable and repeated with fixed contrasts (see Table~\ref{table:binariesdetected2}).}
\tablenotetext{c}{For this data set, the separation was well-determined but the contrast was poorly constrained due to degeneracy. The fit was repeated using the separation determined from the previous night ($\rho = 35.62 \pm 0.51$\,mas) as a prior, giving a revised solution of $C=3.18 \pm 1.59$, $\rho = 35.90 \pm 0.47$\,mas and $\theta=223.09 \pm 1.90$\,deg. The latter translates to a mass of $M_2 = 0.55 \pm 0.08$\,\ms \ according to the NextGen isochrones of \cite{1998A&A...337..403B}. }
\tablenotetext{d}{These data were taken with imaging, rather than aperture-masking, with the astrometry derived from the image auto-correlations. }
\end{deluxetable*}
\clearpage
\end{landscape}

\begin{deluxetable}{lclcccccc}
\vspace{1.5cm}
\tabletypesize{\scriptsize}
\tablewidth{0pt}
\tablecaption{Revised Fits For Degenerate Solutions \label{table:binariesdetected2}}
\tablehead{
\multicolumn{5}{c}{} &
\multicolumn{1}{c}{Prior} &
\multicolumn{3}{c}{Refitted Solution} \\
\colhead{Primary} & 
\colhead{Instrument} & 
\colhead{Date} &
\colhead{MJD} &
\colhead{Filter} & 
\colhead{$C$} &
\colhead{$C$} &
\colhead{$\rho$ (mas)} & 
\colhead{$\theta$ (deg)}
}
\startdata
HIP14807        & PHARO & 2007 Nov 29 & 54433.1  & Ks      &  $3.00 \pm 0.06$    &   $3.00 \pm 0.06$  &  $45.43 \pm 1.15$  &  $248.08 \pm 2.30$    \\
HD16760        & NIRC2  & 2009 Nov 20 & 55155.3  & Kcont   & $13.19 \pm 0.94$    &  $14.05 \pm 0.92$  &  $25.80 \pm 1.03$  &  $287.99 \pm 1.89$ \\
HD113449         & PHARO & 2007 Jun 01 & 54252.1  & CH4s   &  $4.62 \pm 0.02$    &   $4.62 \pm 0.02$  &  $21.97 \pm 0.73$  &  $179.96 \pm 2.84$ 
\enddata
\end{deluxetable}

\begin{deluxetable}{lccc}
\vspace{1.2cm}
\tablewidth{0pt}
\tablecaption{System parameters for binaries with orbits  \label{table:systemparameters}}
\tablehead{
\colhead{}  &
\colhead{HD\,16760} &
\colhead{HD\,113449} &
\colhead{HD\,160934} }
\startdata
$P$ (days)            & $466.5 \pm 0.4$   &  $216.9 \pm 0.2$    &   $3764.0 \pm 12.4$ \\            
$T_0$ (MJD)           & $53336.5 \pm 3$   &  $53410.5 \pm 1$    &   $52389.5 \pm 64$ \\          
$e$                   & $0.084 \pm 0.003$ &  $0.300 \pm 0.005$  &   $0.636 \pm 0.020$ \\           
$i$ (deg)             & $2.6 \pm 0.5$     &  $57.5 \pm 1.5$     &   $82.3 \pm 0.8$ \\   
$a$ (mas)             & $25.5 \pm 2.8$    &  $33.7 \pm 0.4$     &   $152.5 \pm 4.7$ \\                    
$a$ (AU)              & $1.16 \pm 0.18$   &  $0.73 \pm 0.02$    &   $5.05 \pm 0.37$    \\
$\Omega$ (deg)        & $86.9 \pm 1.1$    &  $201.8 \pm 1.6$    &   $266.7 \pm 0.6$ \\                        
$\omega$ (deg)        & $243 \pm 2$       &  $114.5 \pm 0.5$    &   $216.0 \pm 3.1$ \\           
$M_{\rm{total}}$ (\ms)  & $0.96 \pm 0.44$  &  $1.10 \pm 0.09$     &   $1.21 \pm 0.27$ \\
\enddata
\end{deluxetable}



\begin{deluxetable}{lcccc}
\vspace{1.2cm}
\tablewidth{0pt}
\tablecaption{HD\,160934B Astrometry  \label{table:hd160934_allastrom}}
\tablehead{
\colhead{Date} &
\colhead{MJD} &
\colhead{$\rho$ (mas)} & 
\colhead{$\theta$ (deg)} & 
\colhead{Source}}
\startdata
1998 Jun 30 & 50994   & $155 \pm 1$     & $275.5 \pm 0.2$  & \cite{2007AA...463..707H} \\
2005 Apr 18 & 53478.9 & $213 \pm 2$     & $268.5 \pm 0.7$  & \cite{2007ApJ...670.1367L} \\
2006 Jul 8  & 53924   & $215 \pm 2$     & $270.9 \pm 0.3$  & \cite{2007AA...463..707H} \\
2006 Sep 17 & 53995.2 & $218 \pm 2$     & $271.3 \pm 0.7$  & \cite{2007ApJ...670.1367L} \\
2008 Jun 23 & 54640.3 & $169.4 \pm 0.3$ & $273.3 \pm 0.1$  & This study \\
2010 Apr 26 & 55312.6 & $68.8 \pm 0.7$  & $290.0 \pm 0.6$  & This study \\
2011 Apr 23 & 55674.6 & $20.0 \pm 0.1$  & $18.43 \pm 0.1$  & This study \\
\enddata
\end{deluxetable}


\begin{deluxetable}{cccc}
\vspace{1.2cm}
\tablewidth{0pt}
\tablecaption{HD\,160934B at fixed parallax  \label{table:systemparameters_hd60934_b}}
\tablehead{
\colhead{$\pi$} &
\colhead{Mass} &
\colhead{$K$} &
\colhead{Model Age} \\
\colhead{(mas)} &
\colhead{(M$_\odot$)} &
\colhead{(mag)} &
\colhead{(Myr)}
}
\startdata
28.2 & 0.603$\pm$0.042 & 5.29$\pm$0.02 & $\ga$100 \\
30.2 & 0.526$\pm$0.037 & 5.44$\pm$0.02 & 100$^{+100}_{-50} $ \\
32.3 & 0.463$\pm$0.032 & 5.59$\pm$0.02 & 55$\pm$10 \\
\enddata
\end{deluxetable}

\clearpage
\LongTables 
\begin{landscape}
\begin{deluxetable*}{lcccccccccccccc}
\tabletypesize{\scriptsize}
\tablewidth{0pt}
\tablecaption{Survey Detection Limits \label{table:detlims}} 
\tablehead{ 
\multicolumn{3}{c}{} & 
\multicolumn{12}{c}{Detection limits for each separation annulus}
\\
\multicolumn{3}{c}{} & 
\multicolumn{1}{c}{\hspace{0.3cm}} &
\multicolumn{5}{c}{$\Delta\,m$ (mag)} &
\multicolumn{1}{c}{\hspace{0.3cm}} &
\multicolumn{5}{c}{$M_2$ ($M_J$)}
\\
\multicolumn{1}{c}{Target} & 
\multicolumn{1}{c}{Instrument} & 
\multicolumn{1}{c}{Filter} & 
\multicolumn{1}{c}{} &
\multicolumn{1}{c}{20--40} & 
\multicolumn{1}{c}{40--80} & 
\multicolumn{1}{c}{80--160} & 
\multicolumn{1}{c}{160--240} & 
\multicolumn{1}{c}{240--320} & 
\multicolumn{1}{c}{} & 
\multicolumn{1}{c}{20--40} & 
\multicolumn{1}{c}{40--80} & 
\multicolumn{1}{c}{80--160} & 
\multicolumn{1}{c}{160--240} & 
\multicolumn{1}{c}{240--320} 
}
\startdata
\multicolumn{10}{l}{AB Dor Targets} \\
\hline \\
PW And   &   PHARO   &   CH4s   &&  2.27  &  4.52  &  4.81  &  4.74  &  4.73  &&  381  &  99  &  91  &  95  &  96  \\ 
HIP\,3589   &   PHARO   &   CH4s   &&  3.23  &  5.31  &  5.60  &  5.54  &  5.57 &&  443  &  125  &  105  &  109  &  107  \\ 
HIP\,5191   &   NIRC2   &   Hcont  & &  4.56  &  4.84  &  4.79  &  4.73  &  4.23 & &  111  &  103  &  96  &  100  &  136  \\ 
HIP\,6276   &   NIRC2   &   Kcont  & &  4.07  &  5.33  &  5.25  &  5.22  &  5.21 & &  141  &  72  &  75  &  76  &  77  \\ 
HIP\,12635   &   NIRC2   &   Kcont  & &  3.85  &  4.93  &  4.88  &  4.81  &  4.83 & &  127  &  73  &  75  &  78  &  77  \\ 
HD\,16760   &   NIRC2   &   Kcont  & &  5.16  &  6.20  &  6.17  &  6.13  &  6.13 & &  85  &  49  &  50  &  50  &  50  \\ 
HIP\,13027   &   PHARO   &   Ks  & &  1.53  &  4.38  &  5.30  &  5.42  &  5.39 & &  606  &  155  &  96  & 89  &  91  \\ 
HD\,19668   &   PHARO   &   Ks   &&  0.960  &  4.05  &  4.99  &  5.09  &  5.03 & &  609  &  146  &  89  &  84  &  87  \\ 
HIP\,14807   &   NIRC2   &   CO  & &  3.98  &  5.20  &  5.12  &  5.09  &  5.08 & &  132  & 70  &  73  &  74  &  74  \\ 
HIP\,14809   &   NIRC2   &   CO  & &  3.84  &  4.95  &  4.91  &  4.84  &  4.80  &&  233  &  114  &  118  &  123  &  126  \\ 
HIP\,16563A   &   PHARO   &   Ks  & &  2.02  &  4.71  &  5.66  &  5.72  &  5.67 & &  442  &  101  &  58  &  56  &  58  \\ 
HIP\,16563B   &   NIRC2   &   CO  & &  4.12  &  5.26  &  5.19  &  5.15  &  5.13 & &  60  &  34  &  36  &  36  &  37  \\ 
HIP\,17695   &   PHARO   &   Ks  & &  0.52  &  3.71  &  4.74  &  4.77  &  4.78  && 275  &  57  &  34  &  34  &  34  \\ 
HIP\,18859   &   PHARO   &   Ks  & &  2.10  &  4.76  &  5.73  &  5.80  &  5.77 & &  632  &  180  &  96  &  104  &  94  \\ 
HIP\,19183   &   PHARO   &   Ks  & &  0.32  &  3.49  &  4.65  &  4.68  &  4.69 & &  955  &  384  &  186  &  183  &  181  \\ 
BD+20 1790   &   PHARO   &   Ks  & &  0.89  &  4.02  &  5.04  &  5.08  &  5.09 & &  532  &  106  &  64  &  63  &  63  \\ 
HD\,89744   &   PHARO   &   CH4s  & &  3.63  &  5.69  &  5.98  &  5.94  &  5.95  &&  627  &  269  &  223  &  228  &  227  \\ 
HIP\,51317   &   PHARO   &   Ks   &&  2.52  &  5.07  &  6.01  &  6.08  &  6.07 & &  100  &  28  &  21  &  17  &  20  \\ 
HD\,92945   &   NIRC2   &   Kcont  & &  4.09  &  5.22  &  5.15  &  5.13  &  5.10  &&  127  &  71  &  73  &  74  &  75  \\ 
GJ\,466   &   PHARO   &   Ks  & &  1.66  &  4.44  &  5.39  &  5.45  &  5.46  &&  398 &  85  &  51  &  49  &  49  \\ 
HD\,113449   &   PHARO   &   CH4s  & &  2.88  &  5.12  &  5.52  &  5.47  &  5.45 & &  325  &  90  &  71  &  73  &  74  \\ 
EK Dra   &   PHARO   &   Ks &  &  0.80  &  3.98  &  4.94  &  4.98  &  4.98 & &  776  &  224  &  121  &  117  &  118  \\ 
HIP\,81084   &   PHARO   &   CH4s  & &  2.23  &  4.54  &  4.94  &  4.90  &  4.90  &&  200  &  58  &  49  &  49  &  49  \\ 
HIP\,82688   &   PHARO   &   CH4s &  &  2.71  &  4.90  &  5.21  &  5.17  &  5.16  &&  520  &  161  &  132  &  136  &  137  \\ 
HD\,160934   &   PHARO   &   H &  &  2.11  &  4.22  &  4.76  &  4.81  &  4.64 & &  309  &  96  &  70  &  68  &  75  \\ 
HIP\,106231   &   PHARO   &   CH4s  & &  3.18  &  5.26  &  5.55  &  5.51  &  5.46  &&  190  &  61  &  53  &  54  &  55  \\ 
HIP\,110526   &   PHARO   &   CH4s  & &  0.67  &  3.76  &  3.85  &  3.85  &  3.72 & &  282  &  61  &  58  &  58  &  62  \\ 
HIP\,113579   &   NIRC2   &   Hcont  & &  5.24  &  5.51  &  5.42  &  5.36  &  5.14 & &  98  &  90  &  96  &  99  &  104  \\ 
HIP\,114066   &   PHARO   &   CH4s  & &  1.82  &  4.19  &  4.51  &  4.50  &  4.41 & &  271  &  74  &  63  &  63  &  66  \\ 
HIP\,115162   &   NIRC2   &   Kcont  & &  4.34  &  5.60  &  5.51  &  5.47  &  5.49  &&  130  &  67  &  71  &  72  &  71  \\ 
HIP\,118008   &   NIRC2   &   Kcont &  &  3.88  &  5.11  &  5.02  &  5.01  &  4.97 & &  129  &  67  &  71  &  72  &  73  \\

\\
\multicolumn{10}{l}{$\beta$ Pic Targets} \\
\hline \\
HR\,9   &   NIRC2   &   Hcont  & &  4.25  &  4.54  &  4.49  &  4.41  &  4.04 & &  134  &  110  &  114  &  120  &  154  \\ 
HIP\,10680   &   NIRC2   &   Kcont &  &  3.80  &  5.15  &  5.05  &  5.00  &  4.97  &&  86  &  30  &  33  &  34  &  35  \\ 
HIP\,11437B   &   NIRC2   &   Kcont  & &  3.71  &  4.91  &  4.79  &  4.78  &  4.79 & &  23  &  16  &  16  &  16  &  16  \\ 
HIP\,11437A   &   NIRC2   &   Kcont  & &  4.27  &  5.42  &  5.30  &  5.30  &  5.31 & &  27  &  17  &  18  &  18  &  18  \\ 
HIP\,12545   &   NIRC2   &   Kcont &  &  3.50  &  4.89  &  4.84  &  4.78  &  4.81 & &  68  &  21  &  22  &  22  &  22  \\ 
51 Eri   &   NIRC2   &   Kcont &   &  5.19  &  6.29  &  6.21  &  6.21  &  6.18  &&  73  &  24  &  26  &  26  &  27  \\ 
HIP\,25486   &   PHARO   &   Ks &  &  1.47  &  4.33  &  5.32  &  5.37  &  5.37 & &  489 & 75  &  35  &  33  &  33  \\ 
GJ\,803   &   NIRC2   &   Hcont &  &  5.46  &  5.75  &  5.68  &  5.64  &  5.36  &&  16  &  15  &  15  &  15  &  16  \\ 
BD$-$17 6128   &   PHARO   &   CH4s  & &  3.20  &  5.57  &  5.57  &  5.56  &  5.49 & &  90  &  20  & 20  &  20  &  20  \\ 
HIP\,112312A   &   NIRC2   &   Kp  & &  4.29  &  5.31  &  5.19  &  5.08  &  5.04 & &  18  &  13  &  13  &  14  &  14  \\ 
HIP\,112312B   &   NIRC2   &   Hcont  & &  3.34  &  3.70  &  3.57  &  3.53  &  3.25 & &  19  &  17  &  18  &  18  &  20  \\

\\
\multicolumn{10}{l}{Her-Lyr Targets} \\
\hline  \\
HD\,166   &   PHARO   &   CH4s &  &  2.74  &  4.88  &  5.13  &  5.10  &  5.08 & &  396  &  126  &  110  &  112  &  113  \\ 
HD\,10008   &   NIRC2   &   Kcont  & &  3.74  &  4.89  &  4.80  &  4.79  &  4.78  &&  213  &  112  &  118  &  118  &  119  \\ 
HD\,233153   &   PHARO   &   Ks &  &  0.54  &  3.73  &  4.73  &  4.80  &  4.80 & &  430  &  94  &  60  &  58  &  58  \\ 
HIP\,37288   &   NIRC2   &   Kcont  & &  4.74  &  5.86  &  5.82  &  5.78  &  5.75  &&  66  &  40  &  41  &  42  &  43  \\ 
HD\,70573   &   PHARO   &   Ks  & &  0.43  &  3.60  &  4.58  &  4.70  &  4.66  &&  688  &  227  &  132  &  123  &  126  \\ 
HIP\,53020   &   NIRC2   &   Kcont  & &  4.16  &  5.34  &  5.26  &  5.21  &  5.21 & &  34  &  20  &  20  &  20  & 20  \\ 
HN\,Peg   &   PHARO   &   CH4s   &&  2.76  &  4.92  &  5.17  &  5.12  &  5.12  &&  497  &  174  &  152  &  156  &  156  \\

\\
\multicolumn{10}{l}{Tuc-Hor Targets} \\
\hline \\
HIP\,9141   &   PHARO   &   Ks  & &  0.66  &  3.86  &  4.84  &  4.90  &  4.88  &&  699  &  113  &  60  &  57  &  58  \\ 
HIP\,30030   &   NIRC2   &   Kcont & &  4.64  &  5.77  &  5.71  &  5.67  &  5.63  &&  83  &  41  &  43  &  44  &  44  \\

\\
\multicolumn{10}{l}{TWA Targets} \\
\hline \\
TWA-21   &   CONICA   &   L${}^\prime$   &&  1.43  &  4.41  &  5.38  &  5.30  &  5.15 & &  163  &  19  &  12  &  13  &  14  \\ 
TWA-6   &   CONICA   &   L${}^\prime$   &&  1.25  &  4.34  &  5.28  &  5.12  &  4.55  &&  118  &  15  &  8  &  10  &  14  \\ 
TWA-7   &   CONICA   &   L${}^\prime$   &&  1.64  &  4.56  &  5.38  &  5.26  &  4.02  && 75  &  12  &  6  &  7  &  16  \\ 
TW Hya   &   CONICA   &   L${}^\prime$  & &  2.64  &  5.28  &  6.28  &  6.20  &  6.13  &&  79  &  14  &  7  &  7  &  7  \\ 
TWA-3   &   NIRC2   &   Kp   &&  2.43  &  3.61  &  3.47  &  3.24  &  2.96  &&  69  &  27  &  31  &  40  &  48  \\ 
TWA-14   &   CONICA   &   L${}^\prime$  & &  0.120  &  3.38  &  4.34  &  4.23  &  3.86 & &  316  &  43  &  19  &  20  &  26  \\ 
TWA-13B   &   NIRC2   &   Kp  & &  3.22  &  4.27  &  4.19  &  4.03  &  4.03 & &  73  &  32  &  36  &  41  &  40  \\ 
TWA-13A   &   CONICA   &   L${}^\prime$ &  &  1.70  &  4.57  &  5.52  &  5.47  &  5.32 & &  150  &  18  &  12  &  12  &  13  \\ 
TWA-8B   &   NIRC2   &   Kp   &&  2.69  &  3.89  &  3.78  &  3.59  &  3.44 & &  19  &  13  &  14  &  15  &  15  \\ 
TWA-8A   &   CONICA   &   L${}^\prime$   &&  1.78  &  4.65  &  5.60  &  5.50  &  5.22 & &  81  &  13  &  6  &  7  &  9  \\ 
TWA-9   &   CONICA   &   L${}^\prime$   &&  1.22  &  4.29  &  5.32  &  5.22  &  4.96  &&  106  &  15  &  7  &  8  &  10  \\ 
TWA-23   &   CONICA   &   L${}^\prime$   &&  1.45  &  4.44  &  5.42  &  5.31  &  5.17  &&  157  &  18  &  11  &  12  &  13  \\ 
TWA-25   &   CONICA   &   L${}^\prime$   &&  1.56  &  4.50  &  5.34  &  5.29  &  4.79  &&  173  &  19  &  14  &  14  &  17  \\ 
TWA-10   &   CONICA   &   L${}^\prime$   &&  0.224  &  3.50  &  4.56  &  4.43  &  4.31  &&  190  &  21  &  13  &  14  &  15  \\ 
TWA-11B   &   CONICA   &   L${}^\prime$  & &  0.736  &  4.05  &  5.07  &  5.01  &  4.85  &&  199  &  19  &  12  &  13  &  14  \\ 
TWA-11A   &   CONICA   &   L${}^\prime$  & &  3.91  &  6.40  &  7.35  &  7.23  &  7.17  &&  160  &  21  &  14  &  15  &  15  
\enddata
\tablecomments{Columns 4--8 report the 99.9\% lower limits for the magnitude difference $\Delta m$ of any undetected companions relative to the target star obtained for each of the five separation annuli, which are quoted in units of milliarcseconds: namely, 20--40\,mas, 40--80\,mas, 80--160\,mas, 160--240\,mas,  and 240--320\,mas (see Section~\ref{sec:generallimits}). Columns 9--13 report the corresponding upper mass limits $M_2$ for any undetected companions. The latter values were calculated using the distance and apparent magnitude of the target listed in Table~\ref{table:sample} to first convert the magnitude differences in columns 4--8 into lower limits for the absolute magnitude of any undetected companion. These were then converted to masses using the appropriate group age listed in Table~\ref{table:mg} with the DUSTY isochrones of \cite{2000ApJ...542..464C}. }
\end{deluxetable*}
\clearpage
\end{landscape}

\begin{deluxetable}{lcccccc}
\tabletypesize{\scriptsize}
\tablewidth{0pt}
\tablecaption{Previous Imaging \label{table:imassumpt}}
\tablehead{
\multicolumn{1}{c}{Target} &
\colhead{Inner Angle} &
\colhead{Filter} &
\colhead{Sensitivity} &
\colhead{Minimum Mass} &
\colhead{Ref.} &
\colhead{Other Ref.} \\
\multicolumn{1}{c}{} &
\multicolumn{1}{c}{(mas)} &
\multicolumn{1}{c}{} &
\multicolumn{1}{c}{($\Delta$mag)} &
\multicolumn{1}{c}{(\mj)} &
\multicolumn{1}{c}{} &
\multicolumn{1}{c}{} 
}
\startdata
PW\,And	&	500	&	CH4s	&	9	&	$\leq 20$	&	L07	&	MZ04, L05, \\
&&&&&& MH09, H10	\\
HIP\,5191	&	400	&	H	&	7	&	38	&	C10	&	$\cdots$	\\
HIP\,6276	&	500	&	Ks	&	7	&	30	&	MH09	&	$\cdots$	\\
HD\,19668	&	500	&	Ks	&	7	&	34	&	MH09	&	$\cdots$	\\
HIP\,16563A	&	500	&	H	&	7	&	36	&	B07	&	$\cdots$	\\
HIP\,17695	&	500	&	CH4s	&	10	&	$\leq 20$	&	L07	&	$\cdots$	\\
HIP\,18859	&	1000	&	L${}^\prime$	&	See footnote	&	$\leq 20$	&	H10	&	L07	\\
BD$+$20 1790	&	500	&	Ks	&	7	&	26	&	MH09	&	L05, H10	\\
HIP\,51317	&	420	&	H	&	See footnote	&	$\leq 20$	&	M05	&	L07	\\
HD\,92945	&	300	&	H	&	7	&	36	&	B07	&	L07	\\
GJ\,466	&	1000	&	K	&	5	&	67	&	MZ04	&	$\cdots$	\\
HD\,113449	&	750	&	CH4s	&	11.5	&	$\leq 20$	&	L07	&	H10	\\
EK Dra	&	300	&	H	&	7	&	47	&	B07	&	MZ04, L07, \\
&&&&&& MH09	\\
HIP\,81084	&	500	&	CH4s	&	9.5	&	$\leq 20$	&	L07	&	$\cdots$	\\
HIP\,82688	&	500	&	Ks	&	7	&	48	&	MH09	&	$\cdots$	\\
HD\,160934	&	500	&	CH4s	&	9.5	&	$\leq 20$	&	L07	&	MZ04, L05,\\
&&&&&&  H07	\\
HIP\,106231	&	500	&	CH4s	&	10.5	&	$\leq 20$	&	L07	&	MZ04, L05	\\
HIP\,113579	&	400	&	Ks	&	7 	&	40	&	C10	&	MH09	\\
HIP\,114066	&	500	&	CH4s	&	11.3	&	$\leq 20$	&	L07	&	$\cdots$	\\
HIP\,118008	&	300	&	H	&	7	&	33	&	B07	&	C10	\\
HR\,9	&	400	&	Kp	&	See footnote	&	$\leq 20$	&	K03	&	$\cdots$	\\
HIP\,12545	&	300	&	H	&	7	&	$\leq 20$	&	B07	&	$\cdots$	\\
51 Eri	&	1000	&	L${}^\prime$	&	See footnote	&	$\leq 20$	&	H10	&	$\cdots$	\\
HIP\,25486	&	500	&	Ks	&	7	&	$\leq 20$	&	MH09	&	K07, L05	\\
GJ\,803	&	200	&	Ks	&	See footnote	&	$\leq 20$	&	M05	&	K03, MZ04,\\
&&&&&&  B07, L07	\\
BD$-$17 6128	&	290	&	Ks	&	See footnote	&	$\leq 20$	&	M05	&	$\cdots$	\\
HIP\,112312A	&	300	&	H	&	7	&	$\leq 20$	&	B07	&	$\cdots$	\\
HD\,166	&	850	&	L${}^\prime$	&	See footnote	&	$\leq 20$	&	H10	&	L07	\\
HD\,10008	&	600	&	CH4s	&	10	&	27	&	L07	&	$\cdots$	\\
HD\,233153	&	1000	&	Ks	&	5	&	54	&	C05	&	$\cdots$	\\
HIP\,37288	&	400	&	H	&	See footnote	&	$\leq 20$	&	M05	&	L07	\\
HD\,70573	&	500	&	Ks	&	7	&	40	&	MH09	&	L05	\\
HIP\,53020	&	500	&	CH4s	&	8.4	&	$\leq 20$	&	L07	&	$\cdots$	\\
HN Peg	&	750	&	CH4s	&	12.2	&	$\leq 20$	&	L07	&	MZ04	\\
HIP\,9141	&	300	&	H	&	7	&	24	&	B07	&	MH09	\\
HIP\,30030	&	300	&	H	&	7	&	27	&	B07	&	MH09	\\
TWA-6	&	320	&	Ks	&	See footnote	&	$\leq 20$	&	M05	&	MZ04, L05	\\
TWA-7	&	400	&	H	&	7	&	$\leq 20$	&	L05	&	MZ04	\\
TW Hya	&	400	&	H	&	7	&	$\leq 20$	&	L05	&	MZ04	\\
TWA-3	&	200	&	K	&	4	&	$\leq 20$	&	W99	&	C10	\\
TWA-14	&	300	&	H	&	7	&	$\leq 20$	&	B07	&	MZ04, C10	\\
TWA-8B	&	100	&	Ks	&	See footnote	&	$\leq 20$	&	M05	&	L05	\\
TWA-8A	&	140	&	Ks	&	See footnote	&	$\leq 20$	&	M05	&	$\cdots$	\\
TWA-9	&	300	&	Ks	&	See footnote	&	$\leq 20$	&	M05	&	$\cdots$	\\
TWA-23	&	400	&	H	&	7	&	$\leq 20$	&	C10	&	$\cdots$	\\
TWA-25	&	300	&	H	&	7	&	$\leq 20$	&	B07	&	C10	\\
TWA-10	&	400	&	H	&	7	&	$\leq 20$	&	L05	&	MZ04	\\
TWA-11B	&	200	&	K	&	4	&	$\leq 20$	&	W99	&	$\cdots$	\\
TWA-11A	&	400	&	H	&	7	&	24	&	C10	&	W99	\\
\enddata
\tablecomments{ \ For legend to the references provided in Columns 6 and 7, see Table~\ref{table:sample} footnotes. The quoted sensitivities are effective contrasts, expressed as a magnitude difference between primary and secondary, achieved by previous imaging observations in the specified filter at the quoted inner angle and beyond. These sensitivities are also expressed as minimum detectable companion masses over the mass range 20--80\,\mj, which were calculated by interpolating the DUSTY isochrones of \cite{2000ApJ...542..464C}, as described in Section~\ref{sec:generallimits}.   However, for a subset of the targets we quote minimum detectable companion masses directly from the imaging paper: specifically, for the targets imaged by M05 and K03, we quote the inner separation limits for detectable $10\,$\mj \ companions provided in their survey detection limit tables, and for the targets imaged by H10, we quote the equivalent limits for 20\,\mj \ companions provided in their detection limit figures. }
\end{deluxetable}

\clearpage
\begin{deluxetable}{rcccc}
\vspace{5cm}
\tabletypesize{\scriptsize}
\tablewidth{0pt}
\tablecaption{$f_u$ values (\%) from MC analysis assuming
  $d\mathcal{N}/da \propto a^{-1}$ \label{table:mcresults}}
\tablehead{ 
\multicolumn{1}{c}{} & 
\multicolumn{2}{c}{Full Sample} & 
\multicolumn{2}{c}{Reduced Sample} \\
\multicolumn{1}{c}{} &
\multicolumn{1}{c}{Best} &
\multicolumn{1}{c}{Upper} &
\multicolumn{1}{c}{With Her-Lyr} &
\multicolumn{1}{c}{No Her-Lyr} 
}
\startdata
\\
\multicolumn{5}{l}{$d\mathcal{N}/de \propto 2e$}  \\
\hline $M$ power law & 11 (18) & 14 (24) & 13 (22) & 14 (22) \\
       $q$ power law & 10 (17) & 13 (21) & 12 (20) & 13 (20) \\
       $q$ log-normal& 9 (14) & 11 (18) & 11 (16) & 12 (17) \\
       $M_2 = 20\,M_{\rm J}$ & 15 (26) & 21 (40) & 17 (32) & 19 (32) \\
       $M_2 = 40\,M_{\rm J}$ & 11 (19) & 14 (24) & 13 (22) & 14 (23) \\
       $M_2 = 60\,M_{\rm J}$ & 9 (15) & 11 (19) & 11 (17) & 12 (18) \\
\\
\multicolumn{5}{l}{$e=0$} \\
\hline $M$ power law  & 12 (17) & 15 (23) & 14 (21) & 15 (21) \\
       $q$ power law  & 11 (16) & 14 (21) & 13 (19) & 14 (19) \\
       $q$ log-normal & 10 (13) & 12 (17) & 11 (16) & 12 (16) \\
       $M_2 = 20\,M_{\rm J}$ & 16 (25) & 23 (38) & 18 (31) & 20 (31) \\
       $M_2 = 40\,M_{\rm J}$ & 12 (18) & 15 (24) & 14 (21) & 15 (22) \\
       $M_2 = 60\,M_{\rm J}$ & 10 (14) & 12 (18) & 11 (17) & 13 (17) \\
\\
\multicolumn{5}{l}{$e=0.9$} \\
\hline $M$ power law  & 11 (19) & 13 (24) & 13 (22) & 14 (23)  \\
       $q$ power law  & 10 (17) & 12 (22) & 12 (20) & 13 (21)  \\
       $q$ log-normal & 9 (14) & 11 (19) & 10 (17) & 12 (18)  \\
       $M_2 = 20\,M_{\rm J}$ & 15 (27) & 20 (40) & 16 (32) & 18 (32) \\
      $M_2 = 40\,M_{\rm J}$  & 11 (20) & 14 (25) & 13 (23) & 14 (23)  \\
      $M_2 = 60\,M_{\rm J}$  & 9 (15) & 11 (20) & 11 (18) & 12 (19)  \\
\\
\enddata
\tablecomments{Column 1 gives the values calculated using the full 67 target sample listed in Table~\ref{table:sample} along with our ``best'' values for the ages and distances given in Tables~\ref{table:mg} and \ref{table:sample}, respectively.  Column 2 is the same as Column 1 except that upper values for the ages and distances were used in the calculations, i.e.~the best values plus the corresponding uncertainties listed in Tables~\ref{table:mg} and \ref{table:sample}. Column 3 is the same as Column 1, except that the 9 targets of less certain moving group membership (HD\,89744, HD\,92945, GJ\,466, EK\,Dra, HIP\,30030, TWA-21, TWA-6, TWA-14, TWA-23) were not included in the calculations (see Section~\ref{sec:sample}). Column 4 is the same as Column 3, except that the Her-Lyr targets are also excluded from the calculations. In all columns, values in parentheses are those values obtained using the aperture masking limits alone, while the values without parentheses are those values obtained when the detection limits from previous imaging surveys were incorporated into the calculations.  } 
\end{deluxetable}


\end{document}